\documentclass[usenatbib]{mnras}

\usepackage{newtxtext,newtxmath}

\usepackage[T1]{fontenc}
\usepackage{ae,aecompl}

\usepackage[utf8]{inputenc}
\usepackage{geometry}
\usepackage{graphicx}
\usepackage{epstopdf}
\usepackage{epsfig}

\usepackage{hyperref}

\usepackage{color}

\usepackage{verbatim}

\title[Tidally heated exomoon with a single hot spot]{Enhanced thermal radiation from a tidally heated exomoon with a single hot spot}

\author[J\"ager et al.]{Zolt\'an  J\"ager Jr.$^{1,2,3}$,
Gyula M Szab\'o$^{1,2,4}$
\\
\\
\\
$^{1}$MTA-ELTE Exoplanet Research Group, Hungary\\
$^{2}$MTA-ELTE Lendület Milky Way Research Group, Hungary\\
$^{3}$Baja Astronomical Observatory of the University of Szeged, Szegedi ut KT766, H-6500 Baja, Hungary\\
$^{4}$ELTE E\"otv\"os Lor\'and University, Gothard Astrophysical Observatory, Szent Imre h. u. 112, 9700 Szombathely, Hungary\\
}

\date{Accepted 2021 October 5. Received 2021 September 13; in original form 2021 June 10}

\pubyear{2021}

\begin{document}
\label{firstpage}
\pagerange{\pageref{firstpage}--\pageref{lastpage}}
\maketitle

\begin{abstract}
   An exomoon on a non perfectly circular orbit experiences tidal heating that is capable to significantly contribute to the thermal 
   brightness of the moon. Here we argue that the thermal heat is unevenly distributed on the moon's surface, 
   the emission of the tidal heat is limited to a few hot spots on the surface. 
   A well-known example is the tidally heated Io. Due to their significantly increased temperature, 
   the hot spots enhance the energy emission in thermal wavelengths. 
   We made simulations using Monte-Carlo method to examine this contribution, 
   and to predict about the possible detectability of such a spotted exomoon. 
   We found that in the case of large, Earth sized companions to jupiters around red dwarf stars exhibit 
   a thermal flux that enables the direct detection of the moon, 
   due to its photometric signal that can exceed $\approx$100 ppm in the most favourable configurations.
\end{abstract}

\begin{keywords}
planets and satellites: detection -- methods: numerical 
\end{keywords}


\section{\bf Introduction}


Exoplanets observation history has been extending to three decades, but no confirmed exomoon has found yet. 
\cite{Teachey-2017} described a candidate Neptune sized exomoon orbiting the Jupiter-sized exoplanet Kepler-1625b, 
which still needs to be confirmed \citep{Heller-2019}.

Indirect methods, such as accounting for a TTV of the planet caused by the gravity of the moon was suggested by \cite{Sartoretti-1999}.
This method has also a wide literature. Also, variations of the photometric effects have been suggested, such as a pattern in the Rossiter-McLaughlin effect, or planet-satellite mutual eclipses in the light curves (see
\citealt{Szabo-2006},
\citealt{Cabrera-2007},
\citealt{Simon-2007},
\citealt{Kipping-2009a, Kipping-2009b},
\citealt{Kipping-2009},
\citealt{Simon-2010},
\citealt{Tusnski-2011},
\citealt{Pal-2012},
\citealt{Simon-2012}).
A unification of the dynamical and the photometric effects lead to the photodynamical description of the effects that trace an exomoon (\citealt{Kipping-2011}, \citealt{Kipping-2012}).

A direct photometrical evidence of the moon transit component in the light curves is considered to be the most trustable method for an exomoon discovery, as it can directly reveal the moon itself.
Mutual moon-to-planet occultations or an eclipse of a moon
was similarly considered to be the strong evidence of a successful detection (\citealt{Sartoretti-1999}). Accounting for the reflected starlight only,  these moons are predicted to be detectable only in configurations where the planet is very close to the star.

In systems where the moon is predominantly heated by tidal interactions rather than by the incident starlight, the contribution of the moon was suggested to be detectable in the thermal infrared \citep{Dobos-tidal}, regardless to the planet's semi-major axis and the incident stellar irradiation. The background of the tidal heating is the planet-moon-star tidal interactions, which have a complex physics (\citealt{Barnes-OBrien}; \citealt{Sasaki-2012}). 
The tidal interactions dissipate the dynamical energy and heat up the moon, at the price of orbital evolution.
In general, the end stages lead either to
the escape of the moon from the system, or the collision of the moon to the planet \citep{Alvarado-palya}. 
Both the planet and the moon experiences a tidal heating, but since the planet is much larger than a moon, the heating of the moon is the dominant thermophysical process. This heat has recently been taken into
account in understanding the physics of exomoons, because this process can convert an
icy world to a habitable environment. Also, it potentiates the observability of the moon itself.

\citet{Reynolds-1987}; \citet{Scharf-2006}; \citet{Heller-2013} found that due to tidal heating, 
a moon can be heated up to high temperatures enough to support a habitable environment even very far from the star. If this heating is very effective, the heated moon can be warm enough to support a detection in an eclipse or a moon-to-planet transit, or in occultation.

\cite{Dobos-tidal} discussed how the thermal emission from a moon is enhanced in the presence of significant tidal heating. If a moon orbits a planet close-in, energy is dissipated in tidal distortions, and it increases the moon's temperature due to inelastic friction. The signal  is around or just within the limit of the current instruments, slightly above 50-100 ppm in the best cases.

In this paper, we argue that in case of a tidally heated exomoon, the temperature map is very uneven, as the tidal heat can plausibly redistributed in a highly inhomogeneous way, leading to significant hot
spots on the surface. We also discuss how the presence of such hot spots enhance the visibility and the possible direct detection of the moon itself.

And the paper organized as follows:
In Section 2 and 3 we describe the details of the models.
In Section 4 we describe simulation preset 
In Section 5 we show the results,
we analyze the results of the simulations, and in 5.5 we examine the effects on brown dwarves.
The conclusion and summary are in Section 6 and 7.

\section{\bf Photometric effects of tidal heating}
\subsection{\bf Homogeneous temperature distribution}

Tidal energy transport is described by \citet{Reynolds-1987} and \citet{Meyer-2007}.
They assumed a fixed $Q$ model, where the dissipation coefficient ($Q$) and the elastic rigidity ($\mu$) are constants, independent on the temperature. In realty, these of course depend on the temperature (see \citealt{Fischer-1990}; \citealt{Moore-2003}; \citealt{Henning-melting}; \citealt{Shoji-2014}) which is the main source of limitations of the fixed $Q$ models.

Within these assumptions, the tidal heated temperature can be calculated as: \cite{Peters-2013}.

\begin{equation}
T_{tidal}^4 = 17.183 \cdot  \sqrt{ \frac{392}{9747} \cdot \frac{(\pi \cdot G)^5}{\sigma^2} } \cdot \frac{R_m^5 \cdot \rho_m^{4.5}}{\mu Q}  \cdot e^2 \cdot \beta^{-7.5}
\label{Ttidal}
\end{equation}

\begin{equation}
\beta = P_m^{2/3} \cdot (\frac{G \cdot \rho_m}{6})^{1/3}
\end{equation}

\begin{equation}
W_{tidal} = A_m \cdot \sigma \cdot T_{tidal}^4
\end{equation}

\noindent where $G$ is the gravitational constant, $\sigma$ is the Stefan-Boltzmann constant, $A_m$ is the surface area of the moon,
$R_m$ is the radius of the moon, $\rho_m$ is the density of the moon, $e$ is the eccentricity, $P_m$ is the orbital period of the moon, 
$\mu$ is the elastic rigidity, and $Q$ is the dissipation function of the moon. 
However this model underestimates the tidal heat of the body (\citealt{Ross-1988}; \citealt{Meyer-2007}).

In the viscoelastic model $\mu$ and $Q$ are depend on the temperature
(\citealt{Segatz-1988},  \citealt{Fischer-1990}, \citealt{Moore-2003}, \citealt{Henning-melting}, \citealt{Dobos-tidal}).
This model gives more realistic values, 
but the calculation need numerical steps and the equilibrium temperature can't be expressed analytically as in the fixed $Q$ model (see \citealt{Dobos-tidal} for the detailed description).

\subsection{\bf Non homogeneous model}

Both the fixed Q and viscoelastic models assume that the tidal heat will be evenly distributed on the surface. 
On the contrary, we see that the surface of Io (the most significant tidally heated moon in the Solar System) is highly inhomogeneous. There ares several hotspots on the surface (\citealt{Kleer-hotspot}, \citealt{Gutierrez-hotspot}, \citealt{Kleeretal-hotspot}, \citealt{Mura-hotspot}), 
where the temperature is much higher (up to 2000 K, \cite{McEwen-1998} then the environment (having a median temperature of 105 K, very close to the equilibrium temperature).

\cite{Kleer-hotspot} found that hot spots that are persistently active at moderate intensities tend to occur at different latitudes/longitudes 
than those that exhibit sudden brightening events.
This suggest that the heat conductivity is not uniform, and the heat dissipated in tidal interactions comes out in these hot areas.

We suggest here that a single-spot model better approximates the realistic appearance of tidally heated moons. This model consists of the solid crust and a (or a few) hot spot(s). This configuration is thermophysically plausible, which can be demonstrated by simple considerations. Here we generally follow the descriptions in \citep{book} to show how the heat is redistributed on the surface. 

The ratio of the convective to conductive heat transfer in a fluid is characterised by the Nusselt number. 
The Nusselt number ($Nu$) is a dimensionless number, and describes the ratio of convective to conductive heat transfer at a boundary in a fluid. It can be calculated as a function of Rayleigh number, Reynolds number, Prandtl number, hence it strongly depends on the inner structure and the composition of the moon.

In assumed exomoons with unknown composition, we can just approximate the heat transport, taking Nusselt number of water, ice and soils into account. Pure water has a Nusselt number between 58--63 in the range of the Rayleigh number between 2.5--4.5$\times 10^9$ \citep{Pavel-geo}. 
This means that the convection is about 60$\times$ more effective heat transporter than the heat conduction. Heat transport in a complex mantle -- similar to what we expect in the case of an exomoon, for example -- can be described also in terms of the Nusselt number, but its exact value depends on many material and physical properties of the body. For an order-of-magnitude estimate, we can still consider a ratio of the overall energy output of convective and conductive processes as $W_{convective}/W_{conductive}\geq 60$ for bodies with a complex composition, that is, similar or greater than in the case of water. This is valid
because the heat conductivity of different soils (0.55--2.7 W/m/K, \citealt{Song-geo}) is
compatible to that of water (0.58) and ice (2.7) at the triple point. Therefore, the soild crust of a moon can be considered as a similarly good insulator as water or ice. Since the crust and an assumed ice layer behaves similarly, the Nusselt number in the interior of an exomoon -- where predominantly ice and silicates are mixed with other materials) will also be similar.

Heat conduction is everywhere possible, and its length scale can be characterised by the depth of the layer (the diurnal depth) which is affected by the daily heat changes due to the changing incident heat (starlight) during a ``day'' of the exomoon. The diurnal layer has a depth of a coupe of centimeters in the case of the ground on the Earth or our Moon, and is plausibly expected to be similar in exomoons. This is the depth scale of the layer from where the heat belonging to the incident and emitted thermal radiation heat can be in equilibrium.

The heat transfer by convection occurs only at parts where
tubular voids are formed within the crust, and the heat convection is much more effective in these tubes than on the surface (the difference is roughly 1.5 orders of magnitude). Therefore, the layer affected by the heat convection will be much deeper than the diurnal layer, and the heat transfer will be much more effective. The tidal heat is produced in the mantle of the exomoon, and therefore, convection is the only process that can carry this heat to the surface. Since convection is a very efficient process, it affects only a small fraction on the surface, and the thermophysical balance will be fully supported. (If the size of the hot spots increase over the equilibrium size, the very effective convection cools the mantle, this leads to the cooling of the hot spots and the mantle together, and the extent of convection, thus the size of the hot spots will reduce to an equilibrium size.)

Since the heat source of the convection and the conduction in the crust is different (tidal and incident heat) and the time scales are also different (due to the value of the Nusselt number), these two modes of heat transfer can be considered as mostly independent, decoupled processes, that lead to a crust having an equilibrium temperature of an irradiated body, and few hot spots, having enough temperature to emit all the tidal heat on their surface. The above considerations show that the hot spot model is plausible.

\section{\bf Model}
\subsection{\bf Bolometric energy}

We account for a non-homogeneous temperature distribution on the surface due to
inner convections, cryovulcanism, surface renewal; which forms hot spots.  
The whole tidal energy $W_{tidal}$ is emitted via cryovulcanic activity and warm spots with $T_s$ spot temperature.
The majority of the surface considerated as an excellent thermal insulator, and its temperature is $T_{BB}$ (BB stands for black body).
This is consistent with that, the thermal conductor timescale of the crust is much bigger then the convective transport timescale of the magma below it.

\begin{figure*}
\begin{center}
\psfig{figure=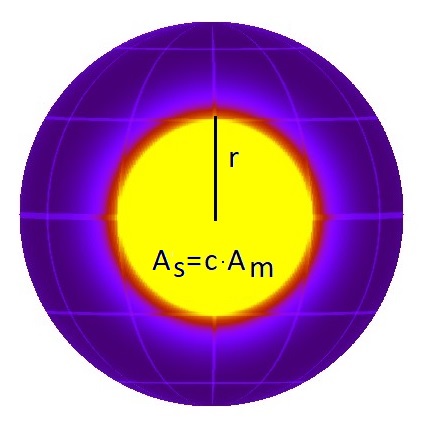,width=0.8\columnwidth} 
\psfig{figure=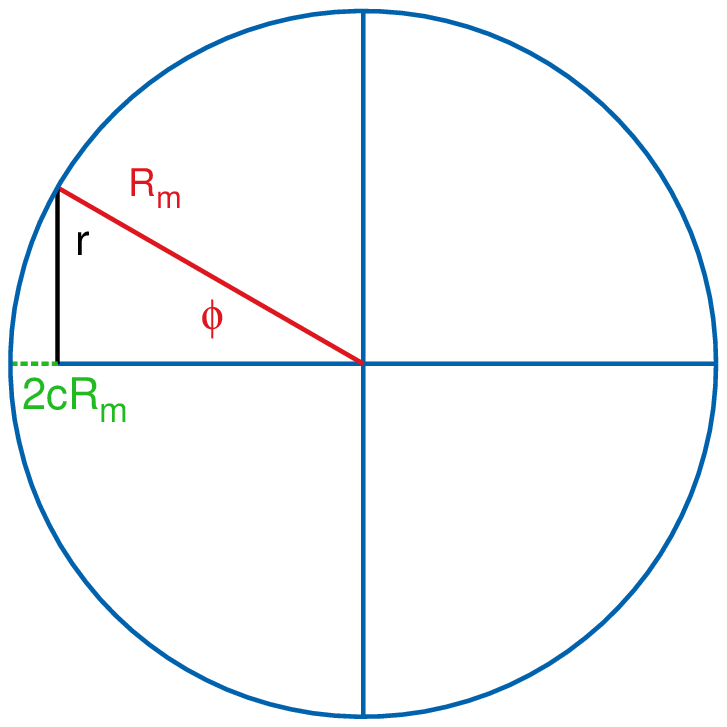,width=1.2\columnwidth}     
\caption{ Left: The geometrical configuration of the tidally heated moon with the hot spot on the surface,
in its maximal possible celestial projection. $A_m$: Moon surface, $A_s$: Spot surface.
Right: Explanation of the geometrical parameters of the spot. $R_m$: Moon radius, $\phi$: Spot radius.
}
\label{kor}
\end{center}
\end{figure*}

The spot is circular, characterised by a solid angle of $\phi$ and an area of $A_s$.
The overall energy output ($W$) is unchanged, but the heat distribution is non-uniform. 
Therefore the side where the spot is brighter. See Fig. \ref{kor}


\begin{equation}
W = W_{BB} + W_{tidal} = (A_m - A_s ) \cdot \sigma \cdot T_{BB}^4 + A_s \cdot \sigma \cdot T_s^4
\end{equation}


Let us also assume that all tidal energy is released in the spot. We assume that the surface temperature is the equilibrium temperature with the stars radiation:

\begin{equation}
T_{BB} = \sqrt[4]{\frac{(1-\alpha) \cdot L}{4 \pi \cdot d^2} \cdot \frac{\alpha}{4 \cdot \sigma} }
\label{tbb}
\end{equation}

where $L$ is the luminosity of the star, $d$ is distance to the star, and $\alpha$ is the albedo.
The moon surface is

\begin{equation}
A_m = 4 \pi \cdot R_m^2
\end{equation}

and the surface of the spot has basic geometrical formulae:

\begin{equation}
A_s = c \cdot A_m = c \cdot 4 \pi \cdot R_m^2
\end{equation}

where c is a constant (between 0 and 1) which measures the $h=c \cdot 2R_m$ sagitta (height) of the circular segment,
It can be calculated from $\phi$ as: 

\begin{equation}
c = \frac{A_s}{A_m} =\sin^2{\frac{\phi}{2}}
\label{c(fi)}
\end{equation}



Therefore:

\begin{equation}
T_s = \sqrt[4]{  \frac{c \cdot W_{BB} + W_{tidal}}{A_s \cdot \sigma} } = \sqrt[4]{  T_{BB}^4 + \frac{T_{tidal}^4}{c}  }
\label{Tscalc}
\end{equation}

The non spotted case is the $c=1$, so $T_{balance}^4 = T_{BB}^4 + T_{tidal}^4$, is the homogeneous temperature.
The spot temperature is $T_{tidal}/\sqrt[4]{c}$ if $T_{BB}$ is neglectable. If we use eq. (\ref{c(fi)}):

\begin{equation}
T_s = \sqrt[4]{  T_{BB}^4 + \frac{T_{tidal}^4}{ \sin^2{\frac{\phi}{2}} }  }
\label{hsmodel}
\end{equation}

The spot temperature inverse proportional with the root of spot radius.

To compare the effect of a spot to non spotted case, 
here we define an amplification factor ($\eta$) expressing the ratio of visible fluxes ($\eta=W_{spotted}/W_{homogeneous}$).
This amplification factor is very similar to the beaming factor 
used at other amplification coming from inhomogeneous surface reflection and reemission (e.g. \citealt{Lellouch-beaming}),
but in our case the cause of the inhomogeneity is coming from the internal structure of the moon.
The bolometric amplification from the homogeneous model (which has $T_{balance}^4 = T_{BB}^4 + T_{tidal}^4$ temperature) 
can be calculated as:\\ 
- if $\phi \leq 90^o$




\begin{equation}
\eta := \frac{W_{spotted}}{W_{homogeneous}}
= \frac{ T_s^4(\phi) \cdot S + T_{BB}^4 \cdot (1-S) }{ T_{balance}^4 }
\label{amplification}
\end{equation}

where S is the projected area of the spot:

\begin{equation}
S = 2 \int_{0}^{\phi} \frac{r}{R_m} \cdot dr = 2 \int_{0}^{\phi} \cos(\phi') \sin(\phi') d \phi' = \sin^2(\phi)
\end{equation}

With this:

\begin{equation}
\eta = \frac{ T_s^4(\phi) \cdot \sin^2(\phi) + T_{BB}^4 \cdot \cos^2(\phi) }{ T_{balance}^4 }
\end{equation}

It can be transformed to:

\begin{equation}
\eta = \frac{ 4 \cdot \cos^2(\frac{\phi}{2}) \cdot T_{tidal}^4 + T_{BB}^4 }{ T_{balance}^4 }
\end{equation}


- If $\phi\geq90^o$:

\begin{equation}
\eta = \frac{ T_s^4(\phi) }{ T_{balance}^4 } = \frac{ \sin^{-2}{\frac{\phi}{2}} \cdot T_{tidal}^4 + T_{BB}^4 }{ T_{balance}^4 }
\end{equation}



Lets examine the amplification with fixed $T_{BB}$, so it only depends on the spot radius:\\
- if $T_{BB}=0$ (or $T_{tidal}>>T_{BB}$, the tidal heat is much higher), $\phi \leq 90^o$:



\begin{equation}
\eta = 4 \cdot \cos^2(\frac{\phi}{2})
\end{equation}

- if $T_{BB}=T_{tidal}$, $\phi \leq 90^o$:



\begin{equation}
\eta = 2 \cdot \cos^2(\frac{\phi}{2}) + 0.5
\end{equation}

\begin{figure}
\begin{center}
\psfig{figure=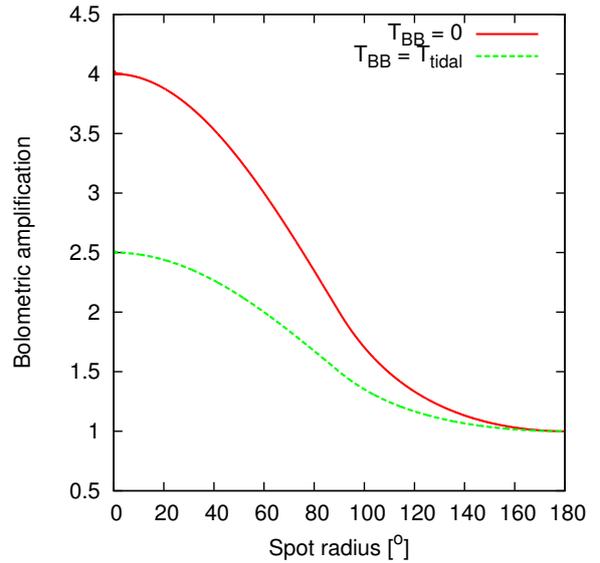,width=\columnwidth}
\caption{ The bolometric amplification ($\eta$) from the homogeneous model, 
i.e. the ratio of the maximal observable flux from the moon without and with a spot, as a function of the $\phi$ spot radius. 
The $T_{BB}=0$ configuration shows the case without incident heating, in the $T_{BB}=T_{tidal}$ case,
the incident and tidal heating are equal.
}
\label{amp}
\end{center}
\end{figure}

The $\eta(\phi)$ functions are plotted on Fig. \ref{amp}. 
As we can see at $\phi=180^o$ the amplification is 1, 
which gives back the homogeneous model, because if $\phi=180^o$ then $c=1$, this means that the whole surface is a spot, and $T_s=T_{balance}$.\\
If $\phi \rightarrow 0$ the amplification is at maximum ($\rightarrow 4$ if $T_{BB}=0$, $\rightarrow 2.5$ if $T_{BB}=T_{Tidal}$). 
However the $\phi=0$ case is not realistic as it implies a spot with zero radius with infinite temperature. 
At $\phi=90^o$, which means the half sphere case, the amplification is 2 (if $T_{BB}=0$) or 1.5 (if $T_{BB}=T_{tidal}$).\\

We can test the prediction of this simple model to the realistic case of Io.
$\sim$30 of Io’s volcanic vents hotter than 700 K, and
in general the higher temperature component is more variable than the lower temperature \citep{McEwen-1998}.
Sulfur boiling in a vacuum cannot get much hotter than about 500 K \citep{Lunine-1985},
so these hot spots are not due to pure sulfur vulcanism. 
Preliminary analyses of the CLR/1MC ratios indicate that temperatures commonly exceed 1300 K, and are sometimes as high as 1500 K or more \citep{McEwen-1998b} on vulcano regions.
When the rapid radiative cooling is considered, these hot-spot observations suggest liquid temperatures hotter than 1700 K \citep{McEwen-1998}.\\
$R_{Io}=1821$ km, $\rho_{Io}=3532$ kg m$^{-3}$, $e= 0.0041$, $\mu Q\approx10^{11}$, $P_{Io}=1.77$ day, 
distance to Sun: 5 AU, we can calculate: $T_{BB}=100$ K, $T_{tidal}=80$ K with fixed Q model using equation (\ref{Ttidal}), so $T_{balance}=110$ K. 
For example $\phi=15^o$ ($c=0.017$) spot radius we get $T_{s}=225$ K with fixed Q,
and $T_{s}=435$ K with viscoelastic. Other values can be seen in Table \ref{tidalT}.
These are quite close to the actual temperature of the areas with constant overheat (\citealt{McEwen-1998}).

\begin{table*}
\begin{center}
\begin{tabular}{ l c c }
\hline

Hot spot size [$^\circ$] & $T_s$ [K] fixed Q & $T_s$ [K] Viscoelastic \\
\hline
180 & 110 & 165 \\ 
 80 & 120 & 200 \\
 30 & 165 & 310 \\
 15 & 225 & 435 \\
  5 & 385 & 750 \\
  2 & 600 & 1200 \\
  1 & 860 & 1680 \\

\hline
\end{tabular}
\caption{Hot spot temperatures on Io calculated from two model. }
\label{tidalT}
\end{center}
\end{table*}


%



\subsection{\bf Corrections from the Nusselt number}

If the Nusselt number is low, then the energy comes out not only in the hotspots, but also in the crust. 
This ratio can be described as a function of the Nusselt number: $f(Nu) \approx \frac{Nu}{Nu+1}$. 
If $Nu>>1$, $f(Nu)\approx 1$ and we get the model assuming perfect heat transfer by convection (Eq. \ref{hsmodel}).

If we consider a surface where convection carries only $f(Nu)$ fraction of the tidal heat to the hot spot, we can write:

\begin{equation}
T_s = \sqrt[4]{  \frac{c \cdot W_{BB} + f(Nu) \cdot W_{tidal}}{A_s \cdot \sigma} }
\end{equation}

\begin{equation}
T_s = \sqrt[4]{  T_{BB}^4 + f(Nu) \cdot \frac{T_{tidal}^4}{c}  }
\label{fnu1}
\end{equation}

And the temperature of the unaffected crust of the moon is:

\begin{equation}
T_m = \sqrt[4]{  \frac{(1-c) \cdot W_{BB} + (1-f(Nu)) \cdot W_{tidal}}{(A_m-A_s) \cdot \sigma} }
\end{equation}

\begin{equation}
T_m = \sqrt[4]{  T_{BB}^4 + (1-f(Nu)) \cdot \frac{T_{tidal}^4}{1-c}  }
\label{fnu2}
\end{equation}


In the followings, we assume $f(Nu)\approx 1$ which is satisfactory for our purposes, and gives back Eq. (\ref{Tscalc}). 
In possible further studies accounting on the deeper details of the thermophysics, we suggest taking Eqs \ref{fnu1} and \ref{fnu2} into account.

\subsection{\bf Wavelength dependence}


In this section we examine the  wavelength-dependent contribution of the excess thermal radiation from the single-spotted exomoon.
The S/N ratio of a detection means the amount of the radiation from the moon, related to that of the host star. 
This ratio increases in towards the thermal infrared domain. 
For example let's take a moon with $T_{\rm moon}=750$ K, $T_{\rm star}=5750$ K. The bolometric the flux ratio is $\frac{T_1^4}{T_2^4}=2.89 \cdot 10^{-4}$, while at 10 $\mu$m,
the wavelength-dependent flux ration is 0.05 which is 170 times bigger, offering a much better observability.

The upcoming Extremely Large Telescope (ELT) with its Mid-infrared ELT Imager and Spectrograph (\citealt{METIS}) 
can observe up to 14 $\mu$m. Hereafter, we use the 14$\mu$m wavelength as a reference. The mission of the Atmospheric Remote-sensing Infrared Exoplanet Large-survey (ARIEL, \citealt{Ariel0})
with an oval 1.1 m $\cdot$ 0.7 m Off-axis Cassegrain telescope,
will be able to perform spectroscopy together with high cadence photometry and synthetic spectrophotometry during transit and eclipse of exoplanets, and to observe the phase-curve. The
ARIEL InfraRed Spectrometer (AIRS) (low/medium resolution, R = 30 - 200) covers the mid-IR and thermal IR between 1.95 and 7.8 $\mu$m
(\citealt{Ariel}).

\begin{figure*}
\begin{center}
\psfig{figure=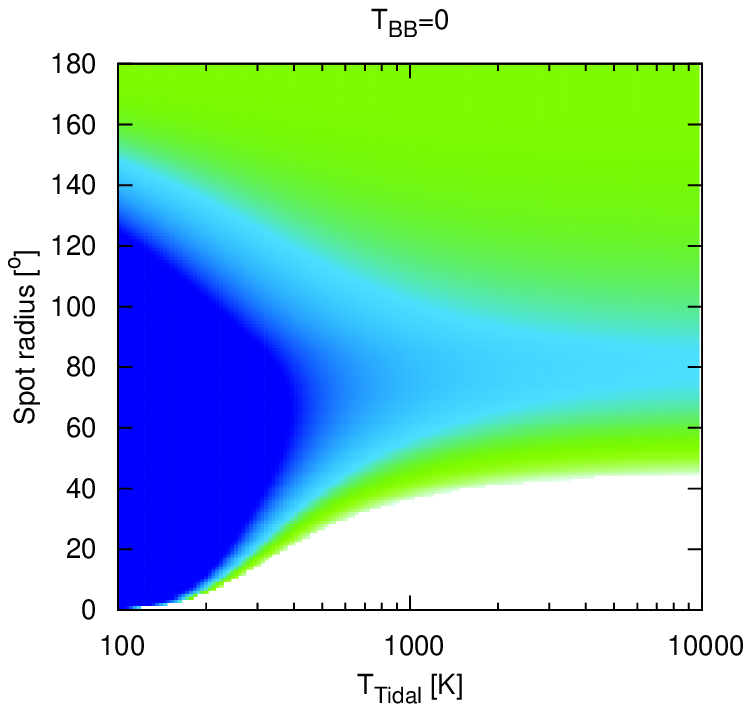,width=0.45\textwidth}
\psfig{figure=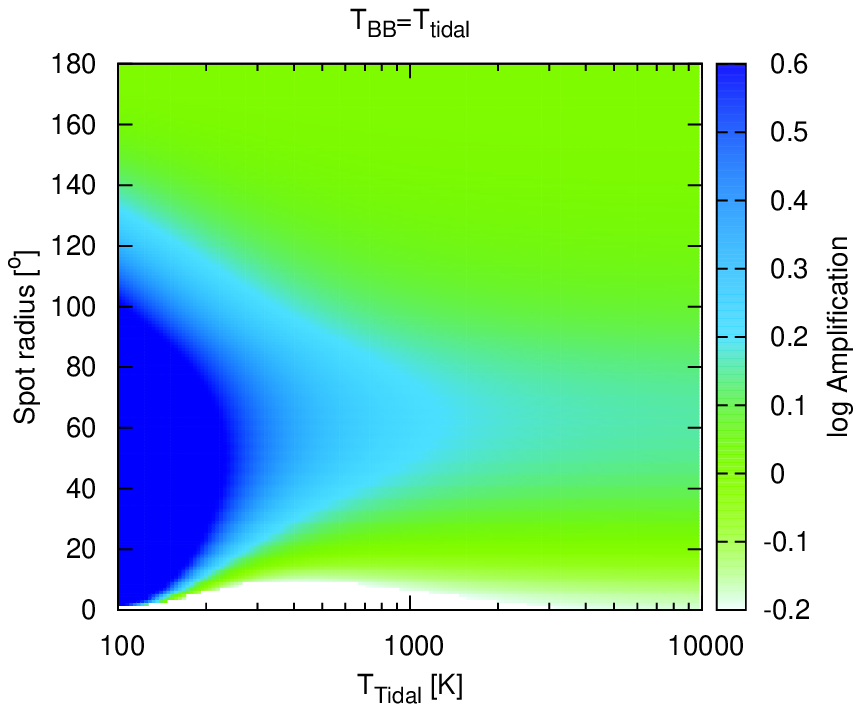,width=0.45\textwidth}
\caption{ The wavelength dependent aplification factor ($\eta$) on 14 $\mu$m, 
as a function of $T_{Tidal}$ (Eq. \ref{Ttidal} for fixed Q model) and Spot radius ($\phi$).
The $T_{BB}=0$ configuration shows the case without incident heating, in the $T_{BB}=T_{tidal}$ case,
the incident and tidal heating are equal.
}
\label{ampNB}
\end{center}
\end{figure*}

In all wavelengths, the amplification will depend on the geometrical and thermal system parameters. 
The contribution of the hot spot alters the observability of the moon, which is expressed in the term of the amplification factor (Eq. \ref{amplification}). 
To illustrate the behaviour of the amplification factor on the important parameters, we consider two special cases in Fig. \ref{ampNB}. The left panel shows the configuration with tidal heating only; the right panel plots the scenario with a tidal heating that is equivalent to the irradiation. 

The uneven thermal emission from the surface leads to an important conclusion: below a certain temperature and spot radius, the amplification becomes smaller then 1, which means the bandpass-dependent signal decreases. 
The reason for this is that if the spot radius becomes small enough, and the temperature increases accordingly (Eq. \ref{amplification}), 
the spectral energy distribution will be shifted to smaller wavelengths.
While the increasing temperature boosts the flux in all wavelengths, the decrease of the area can be a more dominant factor where the changing flux contribution from the spot is small enough.



\subsection{\bf Viscoelastic approximation}


\begin{figure*}
\begin{center}
\psfig{figure=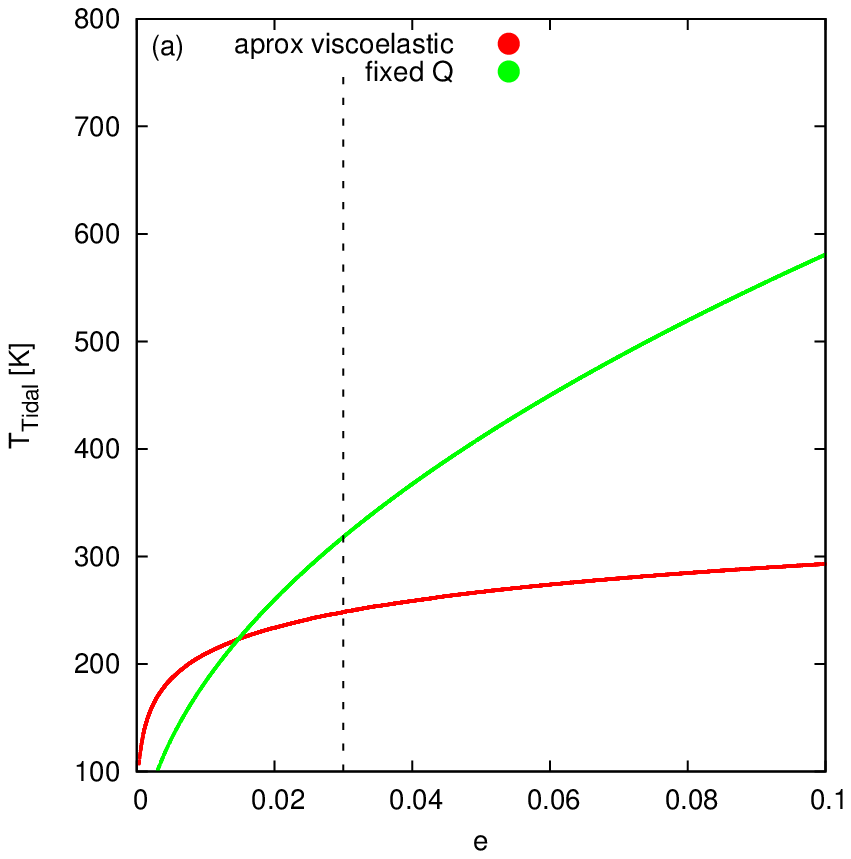,width=\columnwidth}
\psfig{figure=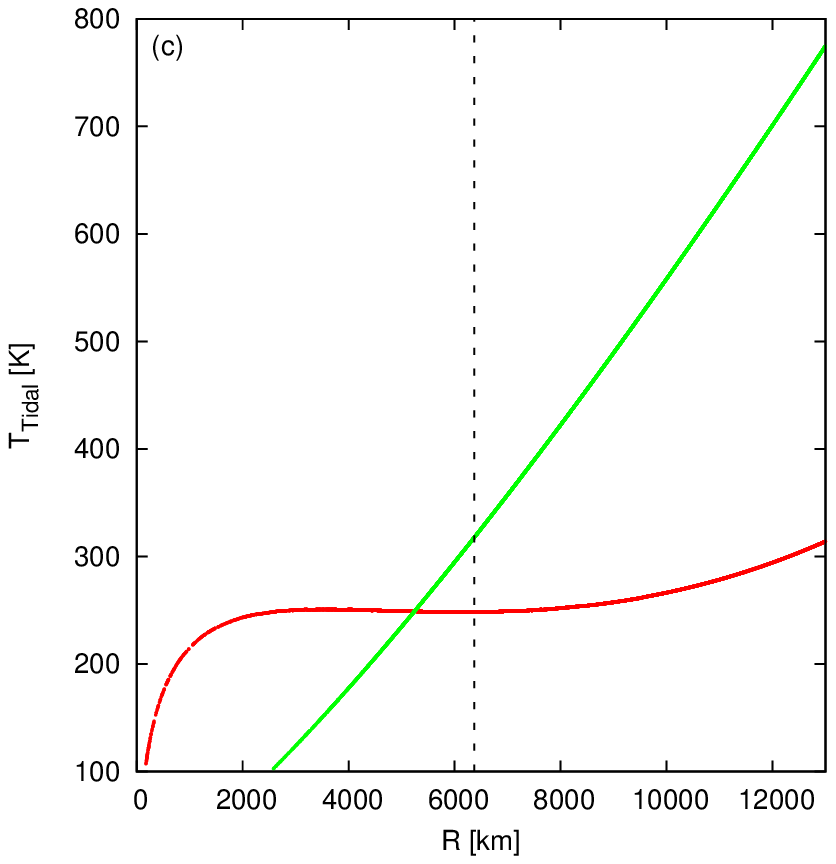,width=\columnwidth}
\psfig{figure=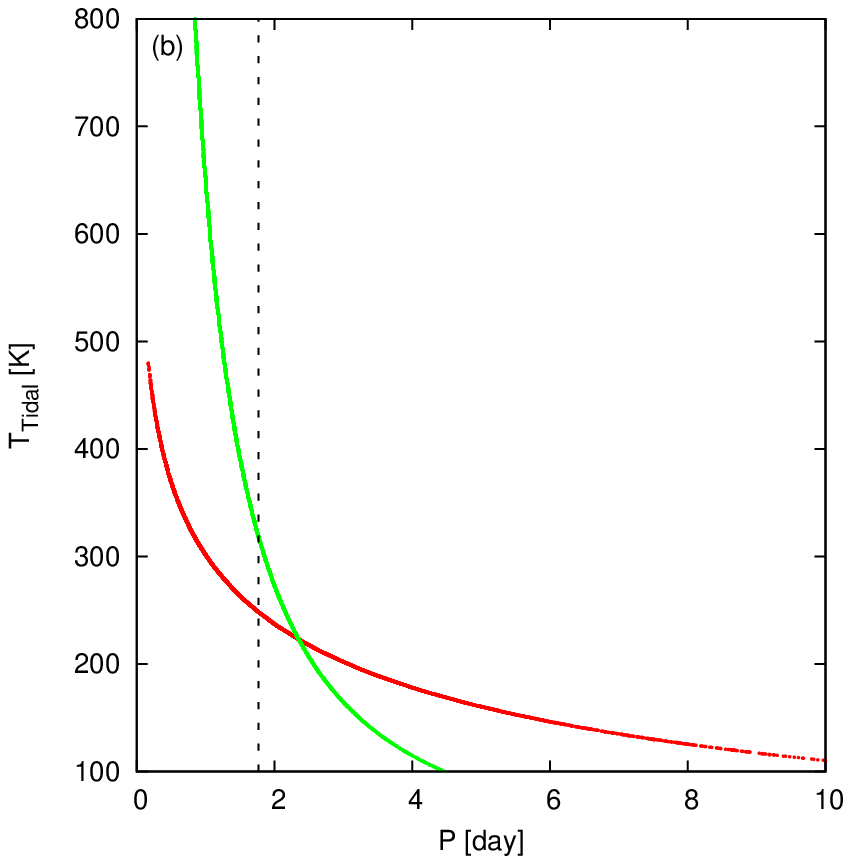,width=\columnwidth}
\caption{ The surface temperature of moons in the viscoelastic and the fixed Q models around the solution belonging to parameters.
The curves show the effect of varing eccentricity ($e$), radius ($R$) and orbital period ($P$) on the surface temperatures.
$\rho=\rho_{earth}$, $R_m=R_{earth}$, $e=0.03$, $P=P_{Io}$, log $\mu Q = 13.5$
}
\label{Thas}
\end{center}
\end{figure*}

The viscoelastic model gives a more realistic model 
where the viscosity and the shear modulus of the body strongly depend on the temperature (see \citealt{Dobos-tidal}),
in contrast to the fixed Q model where $\mu$ and $Q$ are fixed.

We included a analitical model for $\mu$ and $Q$ in the form of a fit to the values in Table 1 of \cite{Dobos-tidal} which is based on \cite{Murray-uQ}.
The result of the fitting is:

\begin{equation}
\log \mu \left( Q (R_m,T) \right) = 
c_1 + c_2 R_m + c_3 R_m^2 + c_4 T,
\end{equation}
where $c_1=5.59 \pm 0.0756$,
$c_2=9.61 \cdot 10^{-7} \pm 2.83 \cdot 10^{-8}$,
$c_3=-4.94 \cdot 10^{-14} \pm 1.97 \cdot 10^{-15}$,
$c_4=0.017 \pm 0.002$.
This approximation gives back the viscoelastic values surprisingly well as can be seen in Fig. \ref{Thas} in present paper, compared to with Figs 5, 6, 7 in \cite{Dobos-tidal}.
The temperature dependency is very well reproduced.
The radius dependency is a bit less accurate as can be seen on $T(R)$ (second panel to Fig 6), but still adequate and results at most 10-15 K difference.
The approximation probably becomes invalid if $T_{tidal}>1600$ K where the body starts to melt (see \citealt{Henning-melting}),
however none of the realistic body reached this temperature.

\section{\bf Simulations Presets}
\label{preset}


We designed computer simulations to decide whatever there is a chance for a positive moon detection within realistic instrumental demands.
The moon-to-star flux ratios were determined in units pf $ppt$, while the total absolute brightness (at 10 pc distance) of the moon only was also calculated as wavelength-dependent AB magnitudes and as bolometric magnitudes.
The parameters of the moon were: $R_m$, $\rho_m$, $e$, $P_m$, and $\phi$ (spot radius). 
The planet parameters were: $M_p$, $\rho_p$ (and so $R_p$ from this) and $d$ distance from the star.
The wavelength was handled as a free parameter, with a special emphasis on the 14 $\mu$m value.
We use MCMC to search for the best moon/star flux ratios (\citealt{Metropolis-mcmc}; \citealt{Hastings-mcmc}; \citealt{Gilks-mcmc}). 
The metropolis ratio in our code was the flux ratio in the sampled wavelength of the moon and the star 
($r=\frac{F_m}{F_{star}}$, accept if $\frac{r_i}{r_{i-1}}>{{\cal U}}[0:1]$).

We had the following constraints for the parameters. The eccentricity was $e<1$, 
and the mass of the planet also need to be much larger then the mass of the moon.
In the solar system, the mass ratio of the gas giants and their moons are at most 10$^{-4}$, which is consistent 
with models of moon formation in a circumplanetary disc, (\citealt{Mosqueira-2003a, Mosqueira-2003b}; 
\citealt{Canup-Ward}; \citealt{Ward-Canup}).
which led to concerns about very large moons of jupiters or ``double planet'' configurations.
However  \cite{Teachey-2017} described a candidate exomoon orbiting the Jupiter-sized exoplanet Kepler-1625b, 
which may have a Neptune sized exomoon, and the mass ratio around was estimated $\sim 5\cdot 10^{-3}$ \citep{Teachey-2018},
and it is possible to be captured \citep{Heller-2018}.
This exomoon however was debated, and needs to be confirmed \citep{Heller-2019}.
The upper limit for the moon size (or mass) of a giant planet is still unknown, and we extended our investigations to a moon size of 2~$R_\oplus$.
We choose the relative mass ratio limit to be $\frac{M_p}{M_m}\geq20$.
The density was independently assigned following a uniform distribution between 2000 to 10000 kg m$^{-3}$ for rocky moons.
Obviously the moon must orbit above the Roche limit of the planet, and the surface of the planet if the planet extends beyond the Roche limit.
The constraints for the orbital period (which contains the moon semi major axis) are the follows:

\begin{equation}
P > \sqrt{\frac{3 \pi}{G \cdot \rho_p}} \cdot (1-e)^{-1.5} \cdot (1+\frac{R_m}{R_p})^{1.5}
\end{equation}

\begin{equation}
P > \sqrt{\frac{9 \pi}{G \cdot \rho_m}} \cdot (1-e)^{-1.5}
\end{equation}

and the moon must be below the Hill radius, so the limit to the orbital period is:

\begin{equation}
P < d^{1.5} \cdot \sqrt{\frac{4 \pi^2}{3 G \cdot M_{star}}} \cdot (1+e)^{-1.5}
\end{equation}

where $d$ is the distance to the star. 

The maximum allowed value of $T_{s}$ was set to 3000 K, close to the hottest lava ocean worlds \citep{Essack-Tlimit}.

\begin{table*}
\begin{center}
\begin{tabular}{ l c c c }
\hline

Parameter                                & $lim$  \\
\hline
Moon excentricity; $e$                   & ]0:1[  \\
Moon orbital period; $P$ [day]           & see Section \ref{preset} \\
Moon mass; $M_{m}$                       & $<M_{P}/20$ \\
Moon radius; $R_{m}$ [km]                & [1000:15000]  \\
Moon density; $\rho_m$ [kg m$^{-3}$]     & [2000:10000] \\
Planet density; $\rho_p$ [kg m$^{-3}$]   & [500:10000] \\
Planet mass; $M_{p}$ [$M_{E}$]           & [50 : 600] \\
Spot Temperature [K]                     & ]0:3000]   \\
Distance to star; $d$ [AU]               & ]0:100]    \\
Spot radius; $\Phi$ [$^o$]               & [3 : 180]   \\
Wavelength; $\lambda$ [$\mu$m]           & [0.5 : 100]  \\

\hline
\end{tabular}
\caption{The input parameter space and their limits of the simulations. }
\label{specfit}
\end{center}
\end{table*}

The planet luminosity comes from the reflection of the star light, 
the reemitted light absorbed by the planet (with $T_{BB}$, see Eq. (\ref{tbb})),
and from the Kelvin-Helmholtz luminosity which is calculated as:
\begin{equation}
L_{KH} = \frac{ G M_p^2 }{ 2 R_p \cdot \tau_{KH} }
\end{equation}
where $\tau_{KH}=30$ Gyr is the Kelvin-Helmholtz timescale \citep{tKH} of the Jupiter which is significantly larger then the age of the system.
So the equilibrium temperature of the planet is:
\begin{equation}
T_p = \sqrt[4]{  \frac{ L_{KH} }{ 4 \pi R_p^2 \cdot \sigma } + T_{BB}^4 }.
\end{equation}

%
%
%
%

\section{\bf Results}

\subsection{\bf Moon-to-system flux rations}

\begin{figure*}
\begin{center}
\psfig{figure=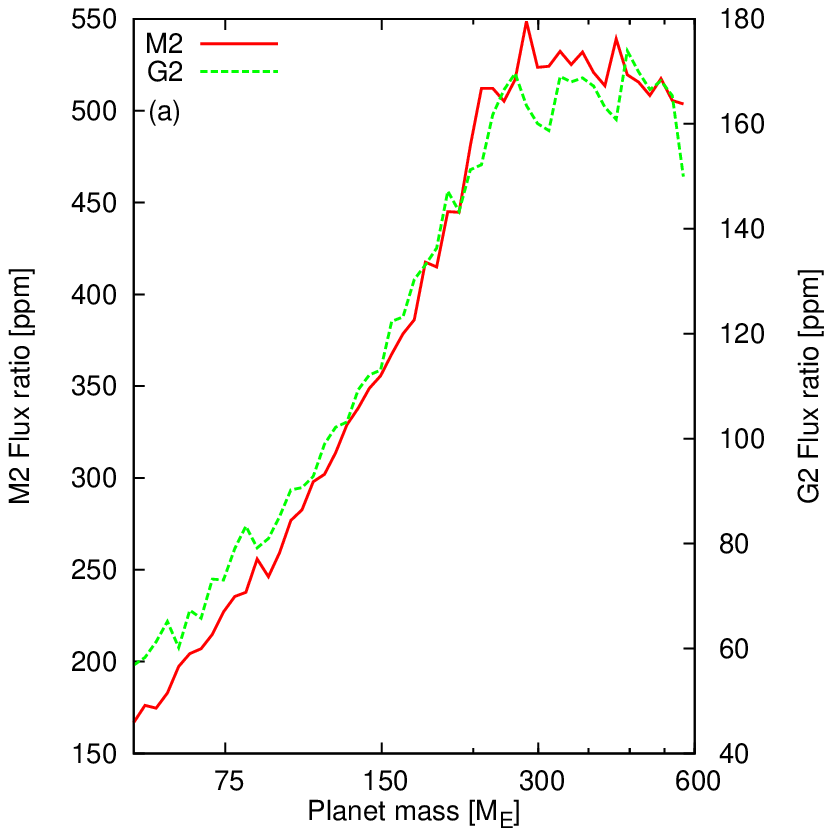,width=0.45\textwidth}
\psfig{figure=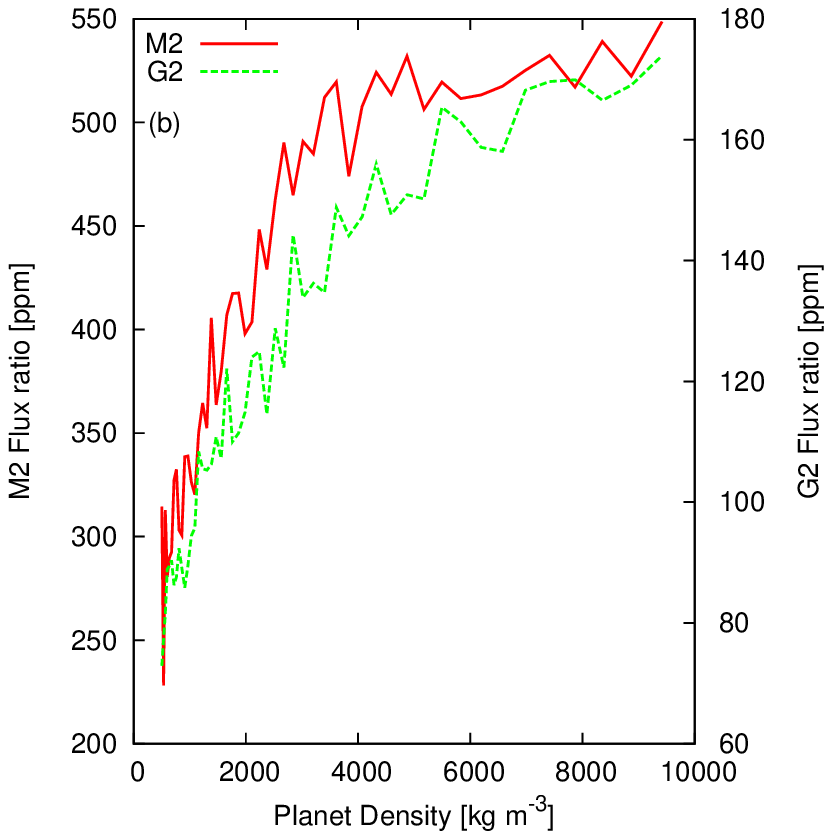,width=0.45\textwidth}
\psfig{figure=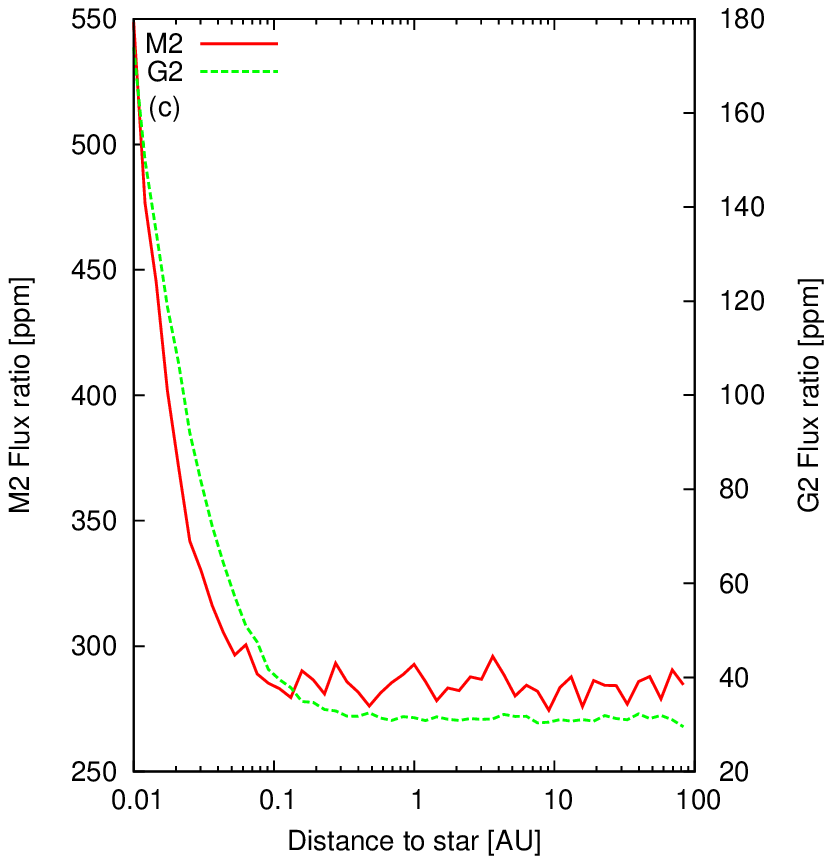,width=0.45\textwidth}
\psfig{figure=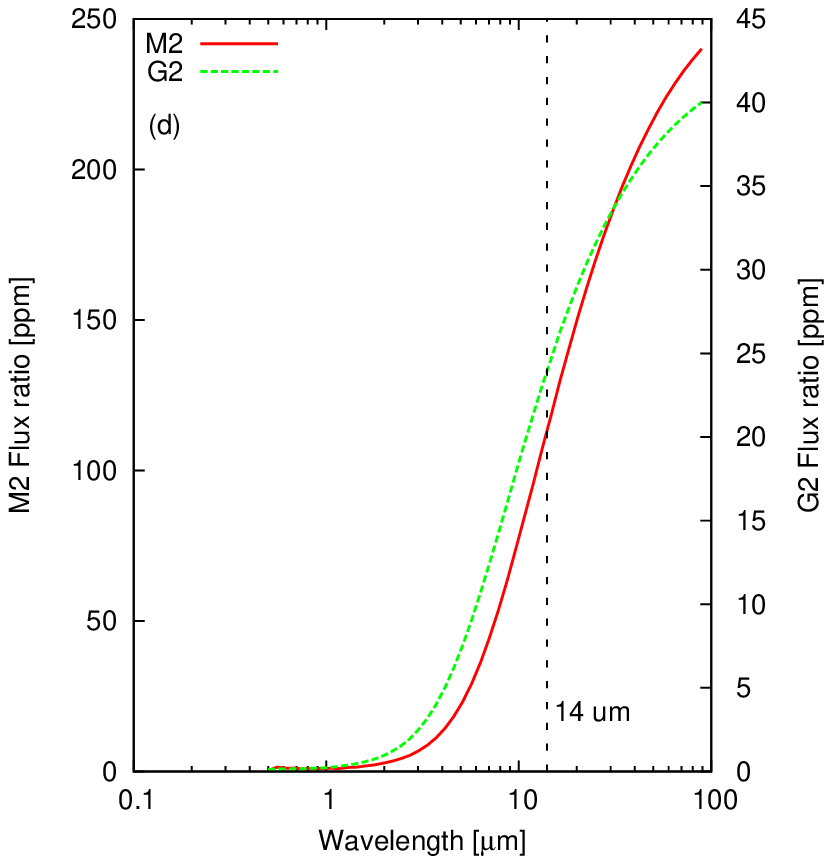,width=0.45\textwidth}
\caption{ Moon-to-star flux ratios, maximum values of the various input parameters in the spotted viscoelastic model. G2 and M2 star. 
Constants of the simulations where:
panels (a)-(c): $\phi=180$, $\rho_m=5500$ kg m$^{-3}$, $\lambda=14$ $\mu$m.
Sampled parameters: $M_p$, $\rho_p$, $d$, $e$, $P$, $R_m$. Their limit can be found in table \ref{specfit}.
Panel (d): Constants of the simulations where: Planet mass: 600 $M_{E}$, planet density: 1300 kg m$^{-3}$, distance to star: 10 AU, 
$R=13000$ km, $\rho=5500$ kg m$^{-3}$, $e=0.2$, $P=$0.25 day. 
$\Phi$ is sampled, and the limit can be found in table \ref{specfit}.
}
\label{margP}
\end{center}
\end{figure*}

\begin{figure*}
\begin{center}
\psfig{figure=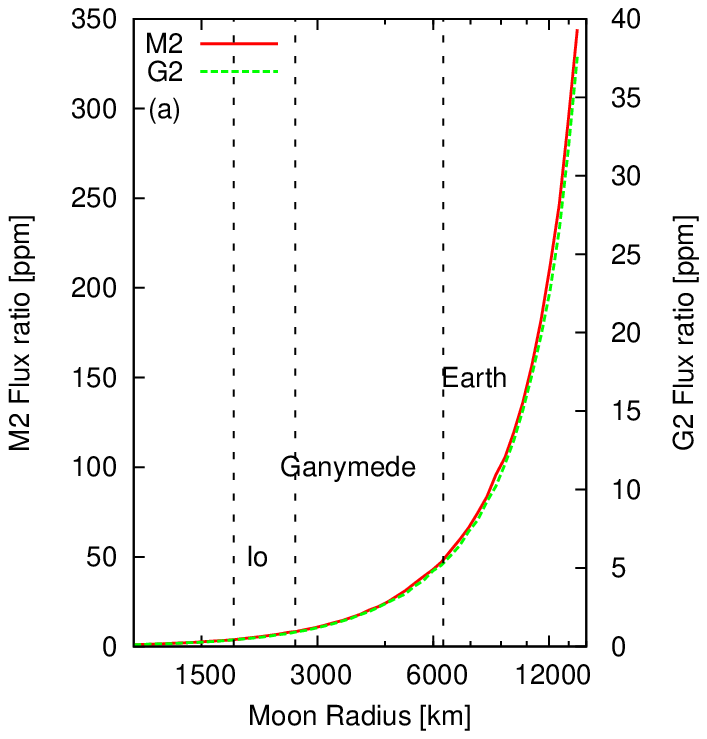,width=0.3\textwidth}
\psfig{figure=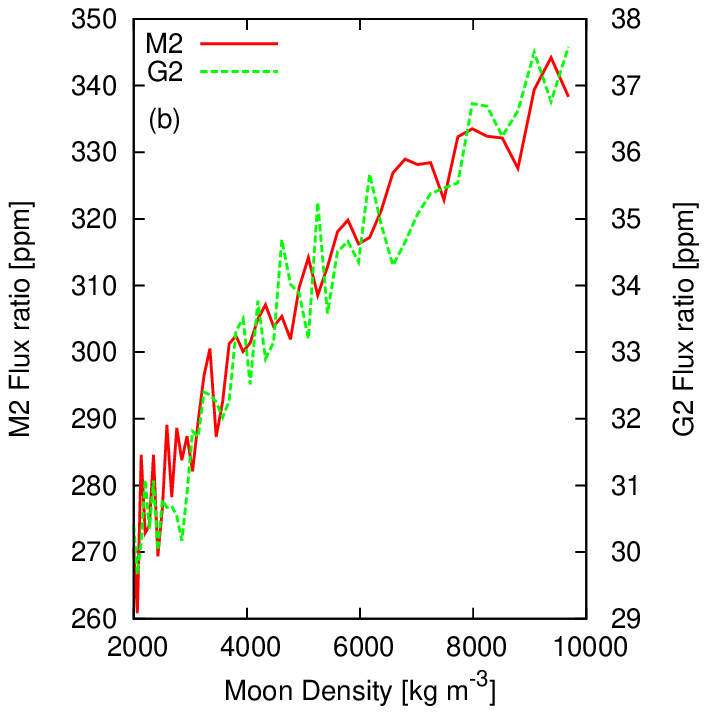,width=0.3\textwidth}
\psfig{figure=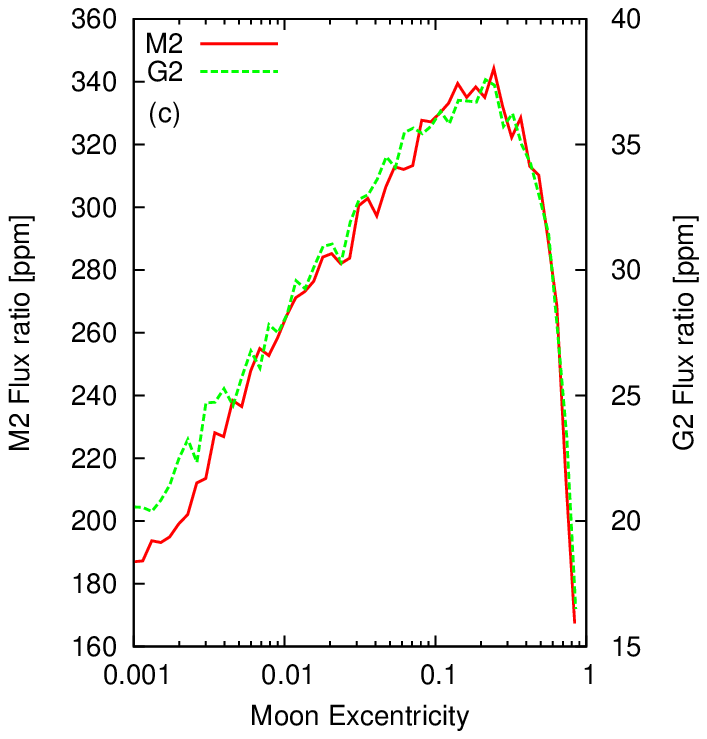,width=0.3\textwidth}
\psfig{figure=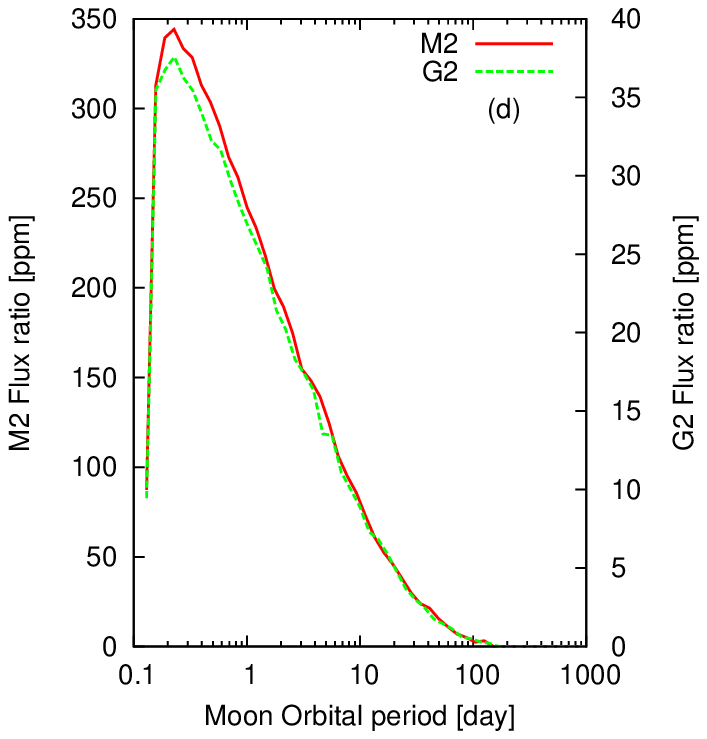,width=0.3\textwidth}
\psfig{figure=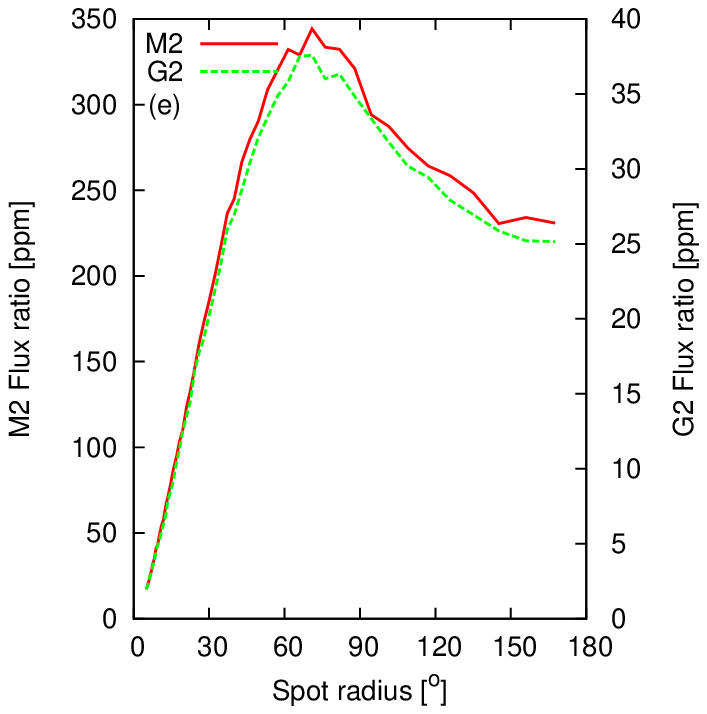,width=0.3\textwidth}
\psfig{figure=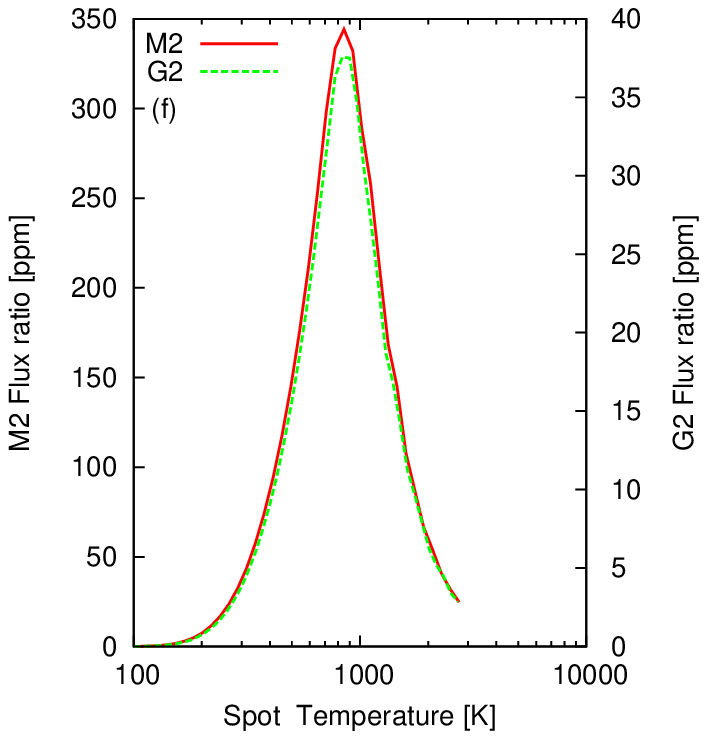,width=0.3\textwidth}
\caption{ Moon-to-star flux ratios, maximum values of the various input parameters in the spotted viscoelastic model. G2 and M2 star. 
Constants of the simulations where:
Planet mass: 600 $M_{E}$, planet density: 1300 kg m$^{-3}$, distance to star: 10 AU, $\lambda=14$ $\mu$m, d$\lambda=0.01\lambda$.
Sampled parameters: $e$, $P$, $R_m$, $\rho_m$, $\Phi$. Their limit can be found in table \ref{specfit}.
}
\label{margK1}
\end{center}
\end{figure*}

Here we describe the results of the simulations.
We calculated the marginalized values for each sampled parameter and plotted them in Figs. \ref{margP}, \ref{margK1}.
In Fig. \ref{margP} the marginalized free parameters are: the mass of the planet (upper left panel), the density of the planet (upper right panel), the star--planet distance (lower left panel), and the wavelength of the observations (lower right panel).

Fig. \ref{margK1} is the same but for marginalised parameters of the planet radius, density, the moon orbital eccentricity (top row), the orbital period of the moon, the $\phi$ extension of the spot and the spot temperature (lower panel).

Since the signal in case of various host stars are proportional to each other, and the shape of the parameters does not depend on the star, two different scales on the two vertical axes could be defined. The left vertical axis shows the flux ratio in case of an M2 host star, the right dependent variable shows a G2 host star.

In an agreement with the intuition, bigger moon have higher fluxes, 
and this is the dominant factor in respect to every other parameters (Fig. \ref{margK1} upper left panel). 
Also the bigger the planet, the higher the moon flux becomes, as the planet allows more massive and bigger moons to orbit (Fig. \ref{margP} upper left panel).
Planet density has less influence, but higher density leads to increased fluxes (Fig. \ref{margP} upper right panel)
because it allows a closer moon, via the Roche-criterion. 
A similar pattern is observed in case of the moon density (Fig. \ref{margK1} upper middle panel), denser moon has higher flux. 
However the radius has far more bigger effect, then the density.
In other words, a mini Neptune is brighter than a super Earth with the same mass, if they orbit a jupiter as its companion.

\begin{figure*}
\begin{center}
\psfig{figure=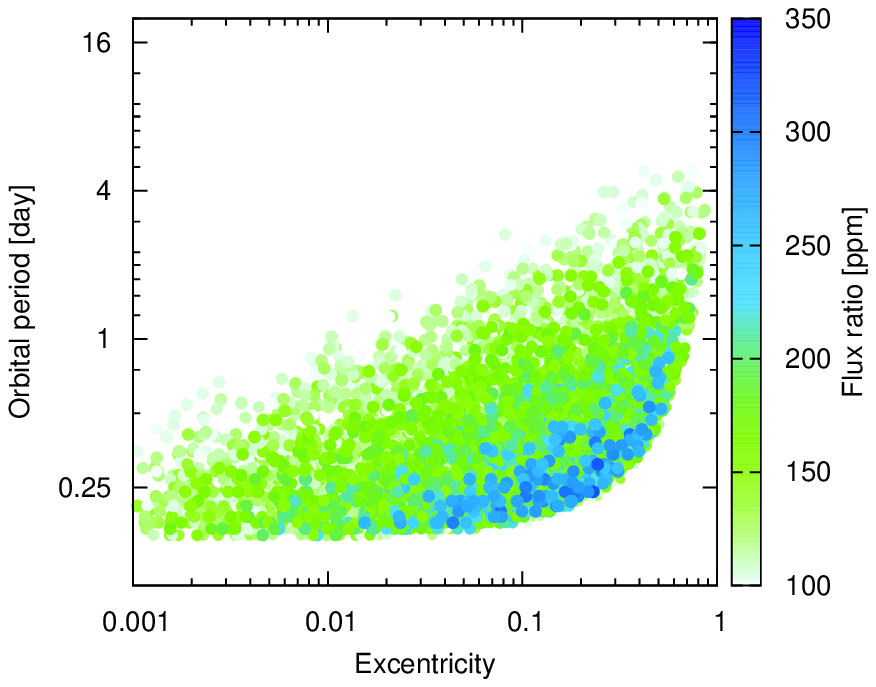,width=\columnwidth}
\psfig{figure=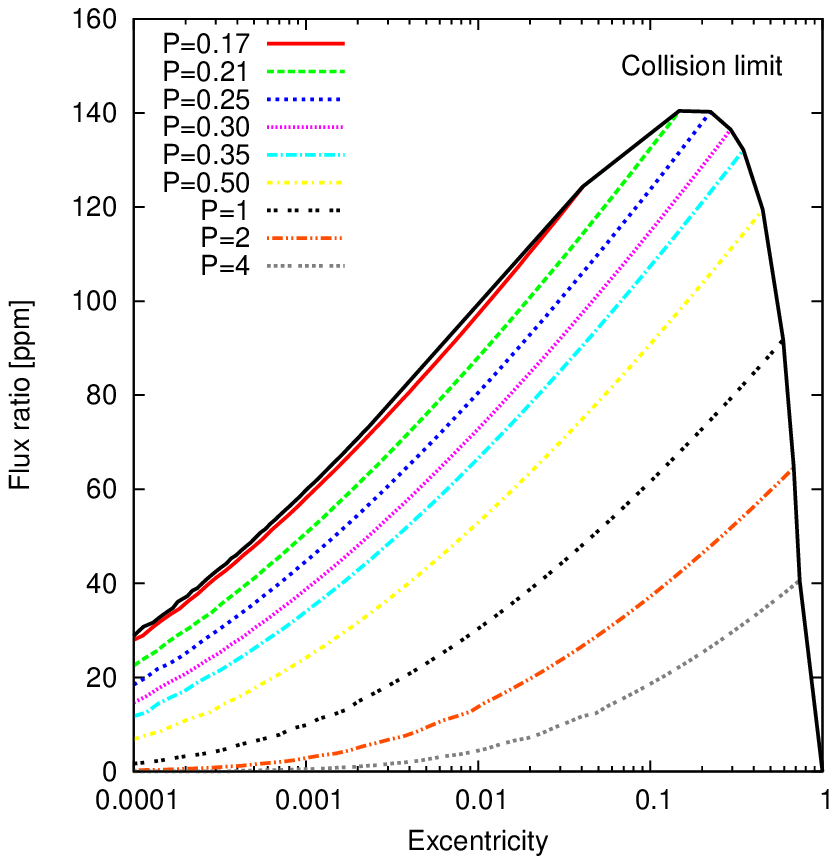,width=\columnwidth}
\caption{ Left: Distribution of moon-to-star flux ratios in the $P$-$e$ space with parameters of: Planet mass: 600 $M_{E}$, planet density: 1300 kg m$^{-3}$, distance to star: 10 AU, $\lambda=14$ $\mu$m, d$\lambda=0.01\lambda$. M2 star.
Sampled parameters: $e$, $P$, $R_m$, $\rho_m$, $\Phi$. Their limit can be found in table \ref{specfit}.
Right: Moon-to-star flux ratios versus $e$ with different Orbital Periods (P, in days).
Note: with high enough $e$ the moon collides with the planet, so bigger $e$ is not possible.
The collision limit of this figure appears as a the non monotonic behaviour in Fig. \ref{margK1} panel (c).
Planet mass: 600 $M_{E}$, planet density: 1300 kg m$^{-3}$, distance to star: 10 AU
The moon is a Super earth: 13000 km, 5500 kg m$^{-3}$.
$\lambda=14$ $\mu$m, d$\lambda=0.01\lambda$, $\Phi=180^{o}$. M2 star.
}
\label{flux}
\end{center}
\end{figure*}

The power of tidal heating is a non monotonical function of the eccentricity, as the maximum flux versus eccentricity has a peak around $e \approx 0.21$  (Fig. \ref{margK1} upper right panel). 
This maximum comes from the fact that the moon must orbit above the Roche limit and the surface of the planet itself, which gives a constrain to the orbital period $P$ which depends on the $e$ (see Section \ref{preset}). 
Bigger moon eccentricity means a closer pericenter, leading to an increased possibility of a collision with the planet. 
This means with low $P$ and high $e$, the moon collides with the planet, and
the orbital period has much greater effect as can be seen in Fig. \ref{flux} left panel.
At a given $P$, higher $e$ always leads to higher heating
(like in a 2 body system; \citealt{Mignard-1980}), but if the $P$ is low, above a given $e$ the collision occur. This phenomenon is demonstrated in Fig. \ref{flux} right panel. 
The second most important parameter (after radius) is the orbital period of the moon (Fig. \ref{margK1} bottom right panel).
The moon orbital period is practically limited by the atmosphere of the planet, and the dynamical stability of the moon in general.

%

\begin{figure}
\psfig{figure=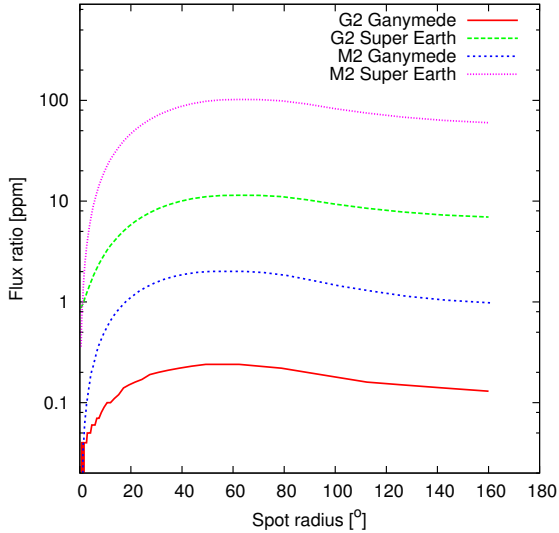,width=0.95\columnwidth}
\caption{ Moon-to-star flux ratios, the values calculated for various moons and stellar types as function of spot radius. 
The other parameters of the simulations are fixed and their values are:
Distance to star: 10 AU, moon orbital periods: 1.77 day (Io orbital period), eccentricity: 0.1, 
planet density: 1300 kg m$^{-3}$, planet mass: 600 $M_E$.
Ganymede: 2631 km, 1942 kg m$^{-3}$.
Super earth: 13000 km, 5500 kg m$^{-3}$.
}
\label{examples}
\end{figure}

If we examine a super Earth exomoon (companion to a jupiter), with 2 Earth radius, Earth density, 1 day moon orbital period ($\sim$ orbital period of the Mimas) and $e=0.1$ in the G2 host scenario, 
we get a signal in the 10 ppm range.
If the planet moon system is close to the star, we get a signal up to 70 ppm (Fig. \ref{margP} bottom left panel),
however in this case the tidal heating is negligible compared
to the heat coming from the star, but moon very close to the star is unlikely.
Taking this system around an M2 star we get a signal around 100 ppm with tidal heating only (far from the star), and taking the incident stellar flux in the habitable zone of the M2 host star into account, too, the signal can reach 125 ppm.
The flux ratio increases toward longer wavelengths (Fig. \ref{margP} bottom right panel).
Small moons produce significantly lesser fluxes as can be seen on Fig. \ref{examples} bottom right panel, and Fig. \ref{margK1} upper left panel.
While a Super Earth around the M2 host provides 100 ppm contribution, 
an Earth sized moon has 10 ppm, and
a Ganymede sized moon only has 1 ppm
in the model system.

\subsection{\bf Moon-to-planet flux ratios}

\begin{figure}
\psfig{figure=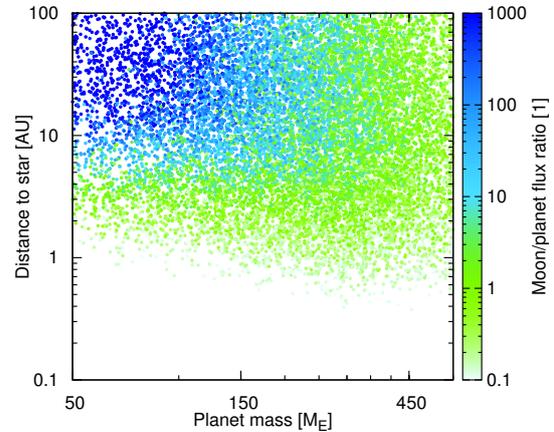,width=0.95\columnwidth}
\caption{ Moon-to-planet flux ratio in the $d$ - $M_p$ space. G2 star.
The fixed parameters of the simulations are:
Wavelength 14 $\mu$m. Spot radius: 180$^{\circ}$, moon density: 5500 kg m$^{-3}$.
Sampled parameters: $M_p$, $\rho_p$, $d$, $e$, $P$, $R_m$. Their limit can be found in table \ref{specfit}.
}
\label{MBd}
\end{figure}

Here we examine the flux ratio of the moon to the planet, showing how much more difficult is a moon detection than the detection of the planet itself. 

In Fig \ref{MBd}, we plot the expected moon-to-planet flux ratios in the G2 host scenario, as a function of the mass and the semi-major axis of the planet.
The relative flux is positively correlated with the distance to the star (as the stellar heating gets less important, and the tidal heating effects will be emphasized) and a negative correlation with the size of the planet (as it scales the reference value of the planet flux; Fig. \ref{MBd}). An important observation is that if the planet is far from the star, the tidal heating of the moon can well outscore the irradiation to both the planet and the moon, and as a consequence, the moon can be much brighter than the planet. This is an important possibility to the detection of planets far out, where practically their tidally heated moons can only (or much better) be observed.
%
%


\subsection{\bf Effects of the spot size}

We assumed that the tidal heat is emitted in a hot spot, instead of along the whole surface.
The spot size can significantly modify the effective temperature of moon, depending on its size.
In our simulations, the spot radius was selected to be a free parameter, however it is unknown which sizes are realistic.
We see on Io that small and medium size spots ($\sim 5^o - 30^o$) are possible,
but large spots ($\sim 80^o$) or rather huge lava lakes on one side of a moon may also conceivable.
The large spots does not increase significantly the temperature, and has a moderate ($\sim2$) amplification effects.
The small spots can alter significantly the temperatures shifting the energy emission to smaller wavelengths, 
while also having a bigger ($\sim3.5$) amplification effects compared to non spotted cases.

As a consequence, the resulting contribution of the tidally heated moon is a non-monotonical function of the spot size. The largest contribution is expected for a spot with around $\phi=70^{\circ}$ parameter, that is, the spot extends to almost one hemisphere (Fig. \ref{margK1} bottom middle panel and on Fig. \ref{examples} bottom right panel).
The difference of the two ``matematically'' unspotted scenarios, $\phi=0^\circ$ and $\phi=180^\circ$ is that in the $\phi=0^\circ$ case, the tidal heat cannot reach the surface (practically the tidal heating is switched off), and in the $\phi=180^\circ$ case, we get the \cite{Dobos-tidal}.
%
It also worth mentioning that if large spots do not form,
then the detection of the moon is more difficult, especially close to the star.

\subsection{\bf Planet-moon systems very far from the star}

\begin{figure*}
\begin{center}
\psfig{figure=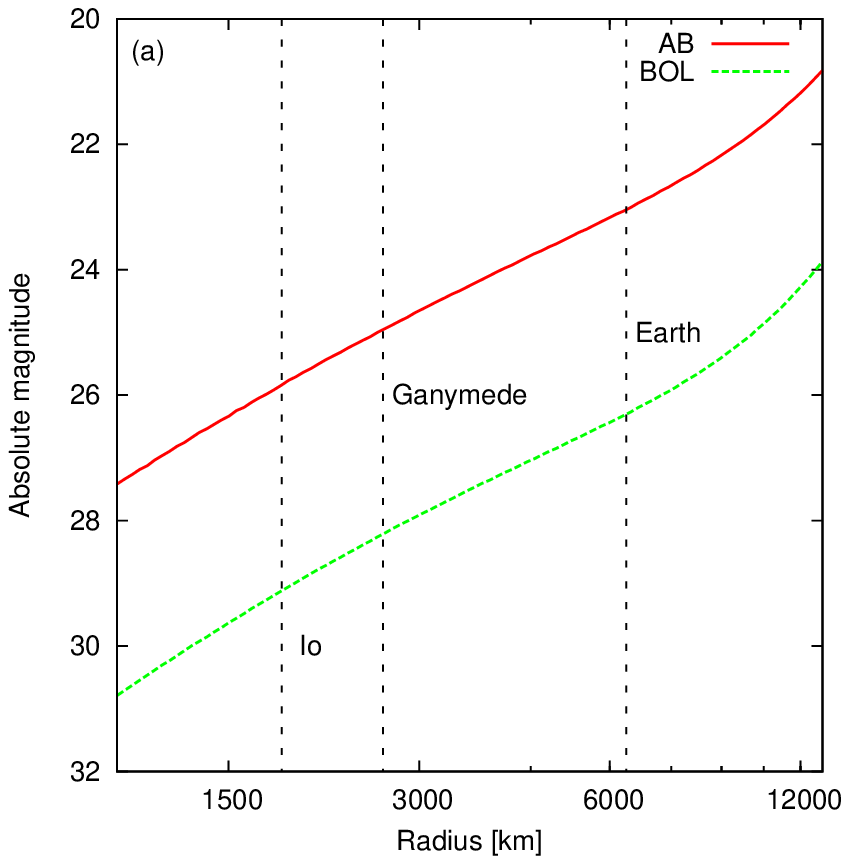,width=0.45\textwidth}
\psfig{figure=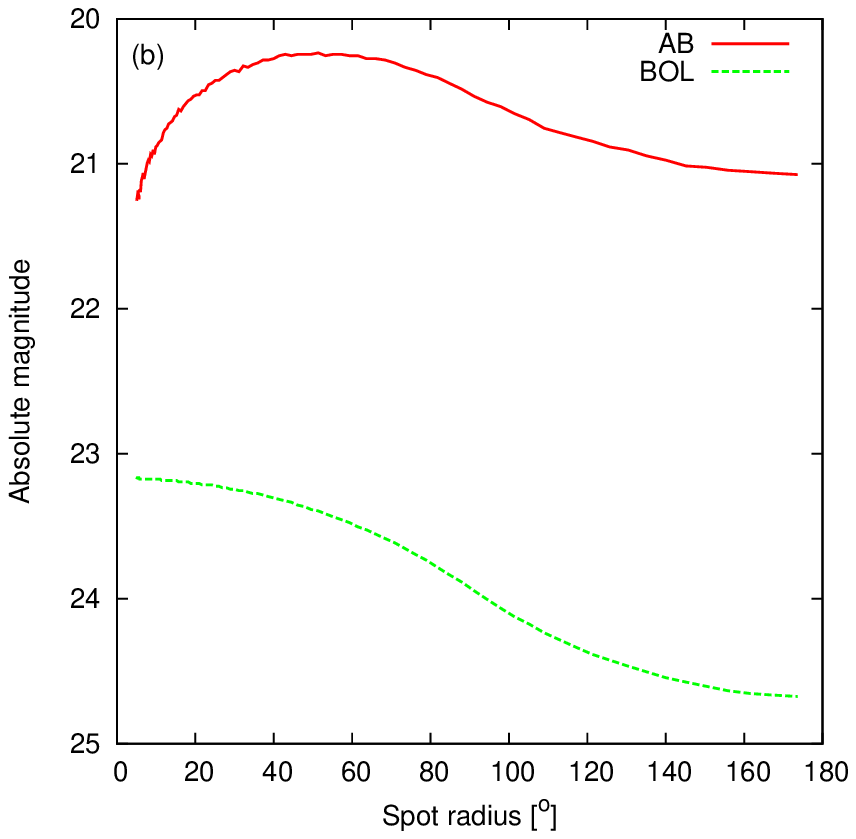,width=0.45\textwidth}
\psfig{figure=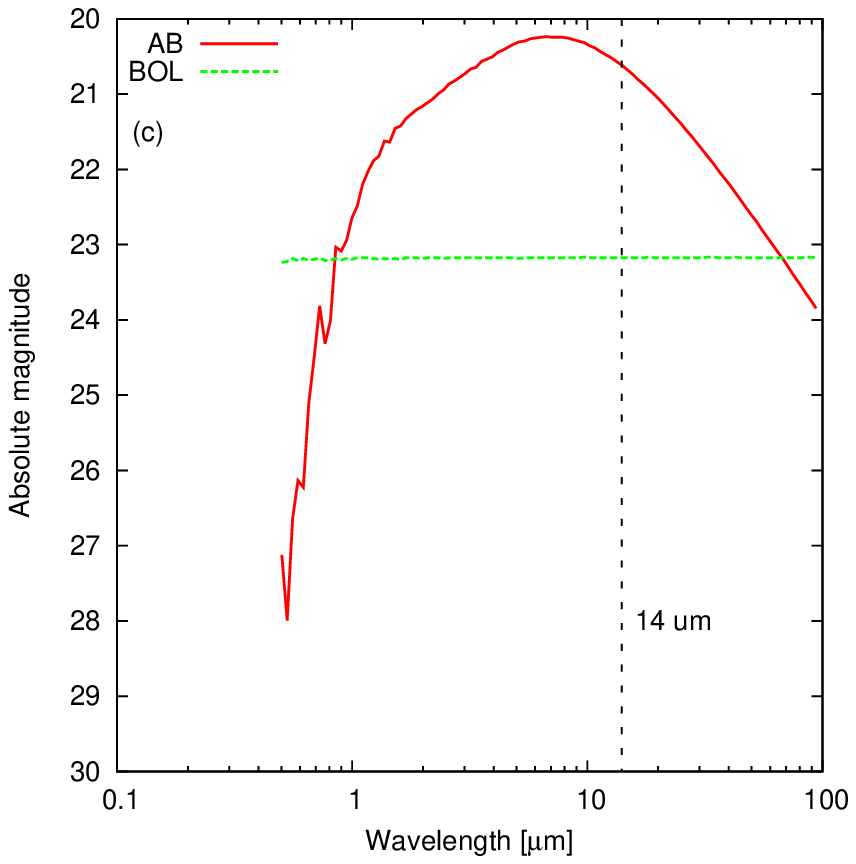,width=0.45\textwidth}
\psfig{figure=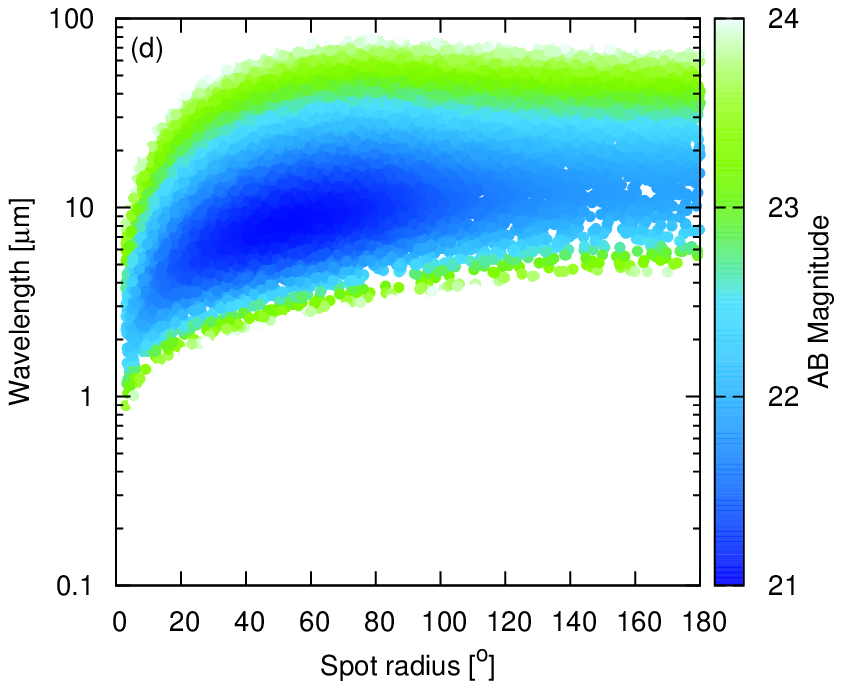,width=0.45\textwidth}
\caption{ Predicted absolute magnitudes of the moon in the spotted viscoelastic model,
maximum values of the various input parameters (panels (a) to (c)), and their distribution of the 
$\Phi$-$\lambda$ space. Fixed parameters are:
$\rho=5500$ kg m$^{-3}$, $e=0.1$, $P=$1 day.
Planet mass: 600 $M_{E}$, planet density: 1300 kg m$^{-3}$. No star. 
Sampled parameters: $R_m$, $\lambda$, $\Phi$. See table \ref{specfit}.
}
\label{margMag}
\end{center}
\end{figure*}


If the moon is very far out from the star and can be imaged with a separation, or if the planet-sized body does not belong to a stellar system at all (rogue planets, \citealt{Bear-rogue}),
the starlight does not directly affect the detection. In this case, the system can be characterised by absolute bolometric magnitudes and absolute AB magnitudes.

Away from the star, the effects of the spots are always positive (Fig. \ref{margMag} upper right panel).
The best detection wavelengths can be around 10--20 $\mu$m (Fig. \ref{margMag} bottom left panel) 
especially if the spot temperature can reach or exceed 1000 K (Fig. \ref{margK1} upper left panel).
The best detection wavelengths however is correlated with the spot radius (Fig. \ref{margMag} bottom right panel).

Super earths around jupiters can reach 21 AB mags and 24 bol mags (Fig. \ref{margMag} upper left panel), which gives then good change to be detected.
An earth-sized moon reach 23 AB mags or 26 bolometric mags, and Ganymedes sized moons has 25 AB mags and 28 bol mags.
This shows that we has fairly good changes to observe moon far from stars in wavelength around 10--14 $\mu$m.

It  is  estimated  that  at least 75\% of systems with giant planets must have experienced planet-planet scattering in the past 
\citep{Raymond-2020}. 
According to planet formation theories, such as the core accretion theory \citep{Ida-2013},  
typical  masses  of  ejected  planets should  be  between  0.3  and  1.0 M$_{Earth}$ \citep{Ma-2016}.
This is not too promising for their detection even via the tidally heated moon, due to the low planet mass. But, if giant planets were also lost from some distant solar systems, and their moon gained significant eccentricity in the scattering scenario, the search for the tidally heated moon is an option that still can be accounted for.

\subsection{\bf A possible interpretations of the NIR excess of Y brown dwarfs}



It is known that there exists two brown dwarf (BD) stars, 
WISEA J053516.87-750024.6 shorthand W0535, and WISEA J182831.08+265037.6 shorthand W1828, 
with a spectral type of Y1 and $\geq$Y2, respectively \citep{Fontanive-2021-Bdwarf}, which exhibit an excess flux in $Y$, $J$ and $H$ bands, 
in respect to the mid-IR flux of these stars. These objects fall $\geq$ 1 magnitude above the brightness-color 
distribution of the known Y1 and Y2 brown dwarfs (\citealt{Leggett-2017-Bdwarf}, \citealt{Kirkpatrick-2019-Bdwarf}, \citealt{Kirkpatrick-2020-Bdwarf}).
Broad-band colors cannot be fit by any of the current BD of models (e.g., \citealt{Beichman-2013-Bdwarf}),
and has a unique spectrum that does not compare well with the known suite of theoretical models (Cushing et al., in prep.).
Both these enigmatic BD stars have a mass in the range of 8-20 Jupiter mass \citep{Leggett-2017-Bdwarf}.

\cite{Tinney-2014-Bdwarf} attempts to interpret the spectrum of W1828 as a a 325K+300K binary. 
\cite{Leggett-2017-Bdwarf} follows this interpretation, and speculate that the object may be an equal-magnitude binary,
however their best fit atmospheric model suggests a young system ($\sim$ 1.5 Gyr), 
which is in contrast with the subsolar metallicity ([M/H]$\approx$-0.5).
They also note that it is hard to reconcile with it being observed at 2 magnitudes above the Cloudless model sequence in MW2 (Fig. 4), 
2.5 magnitudes brighter than  the  H2O  Cloudy  model  sequence,  
and  3.2  magnitudes brighter than W0855 (which is otherwise consistent with the H2O Cloudy model sequence).  
The fits with a binary BD were so unsatisfying that the authors suggested companions in this system.
\cite{Opitz-2016-Bdwarf} used high-resolution adaptive optics imaging in the near-infrared to 
rule out an equal-magnitude binary with a separation >1.9 AU for W1828 and >1 AU for W0535.

\begin{figure*}
\begin{center}
\psfig{figure=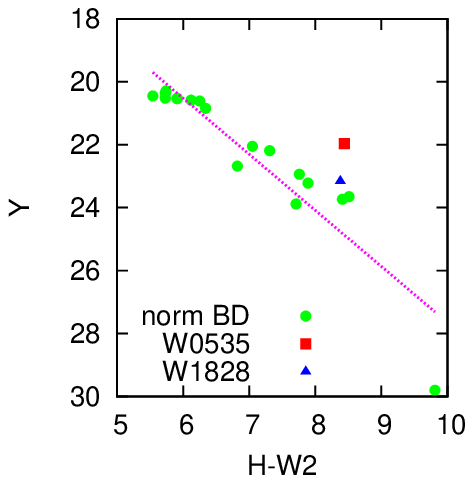,width=0.3\textwidth}
\psfig{figure=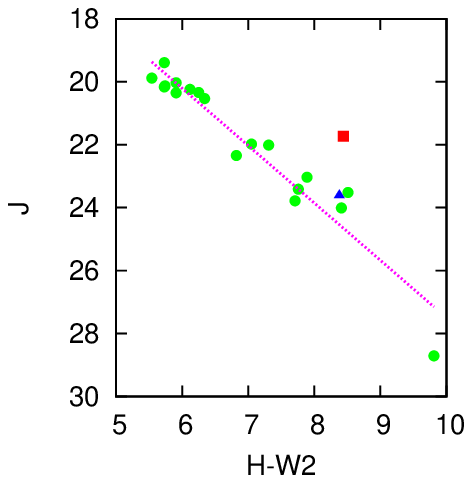,width=0.3\textwidth}
\psfig{figure=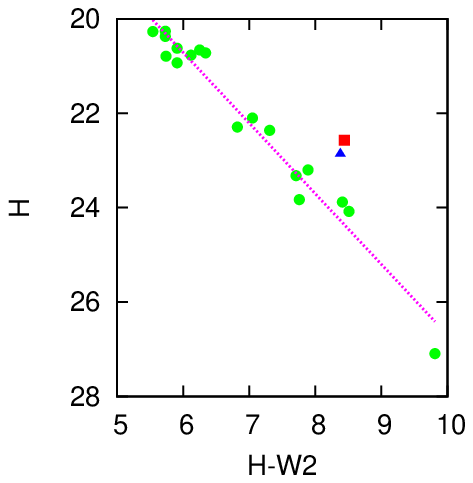,width=0.3\textwidth}
\psfig{figure=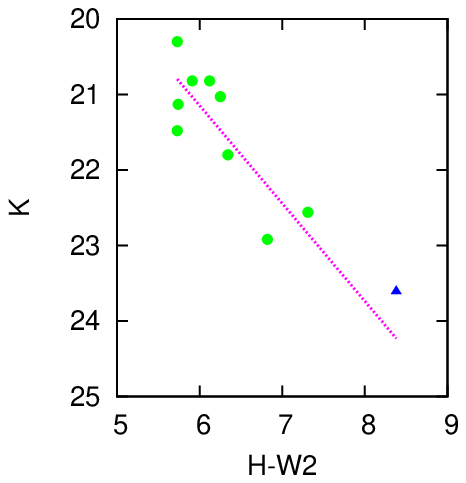,width=0.3\textwidth}
\psfig{figure=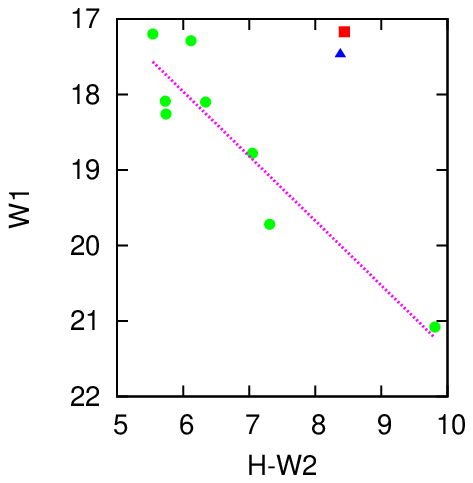,width=0.3\textwidth}
\psfig{figure=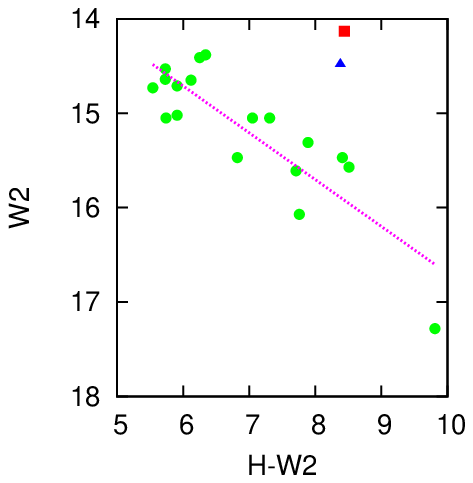,width=0.3\textwidth}
\caption{ H-W2 color magnitude diagram showing the normal brown dwarves, and the two peculiar ones. They show a 1--2 magnitude excess.
Source:  \protect\cite{Tinney-2014-Bdwarf},  \protect\cite{Leggett-2017-Bdwarf},  \protect\cite{Kirkpatrick-2019-Bdwarf},  \protect\cite{Kirkpatrick-2020-Bdwarf} 
}
\label{fits1}
\end{center}
\end{figure*}

Here we examine the possibility that W1828 and W0535 are single BD stars 
with a tidally heated companion (planet or moon -- here we feel the nomenclature has not been solidified by now). 
We consider whether would the excess light in the Y--J--H--K bands, 
would be interpreted as the black body radiation from a hot spot on the assumed companion's surface in our hot spot model.
The observed excess in the $Y$, $J$ and $H$ flux of the W1828 and W0535 spectrum is in the order of 
2.5--1$\times 10^{-22}$~$W\ cm^{-2}\ \mu m^{-1}$, see Fig. \ref{fits1}. 

%
The significant surplus between 1 and 2 $\mu$m could be explained with a 2000 -- 3000 K body.
For first glance, this seems to be a physical for a companion around a BD host, 
but we suggest that it could be achieved by a $\phi=3$--$5^\circ$ sized hot spot on the companion. 
The sufficient size of the companion is the size of Earth,
and it should have an orbital period of $\approx 1$ day ($\sim \leq$ 0.005 AU) if $e\approx0.1$.
The mass ratio of the moon and the Y dwarf is between 10$^{-3}$ -- 10$^{-4}$ which is in the range of
the models of moon formation in a circumplanetary disc (\citealt{Mosqueira-2003a, Mosqueira-2003b}; 
\citealt{Canup-Ward}; \citealt{Ward-Canup}).

\begin{figure*}
\begin{center}
\psfig{figure=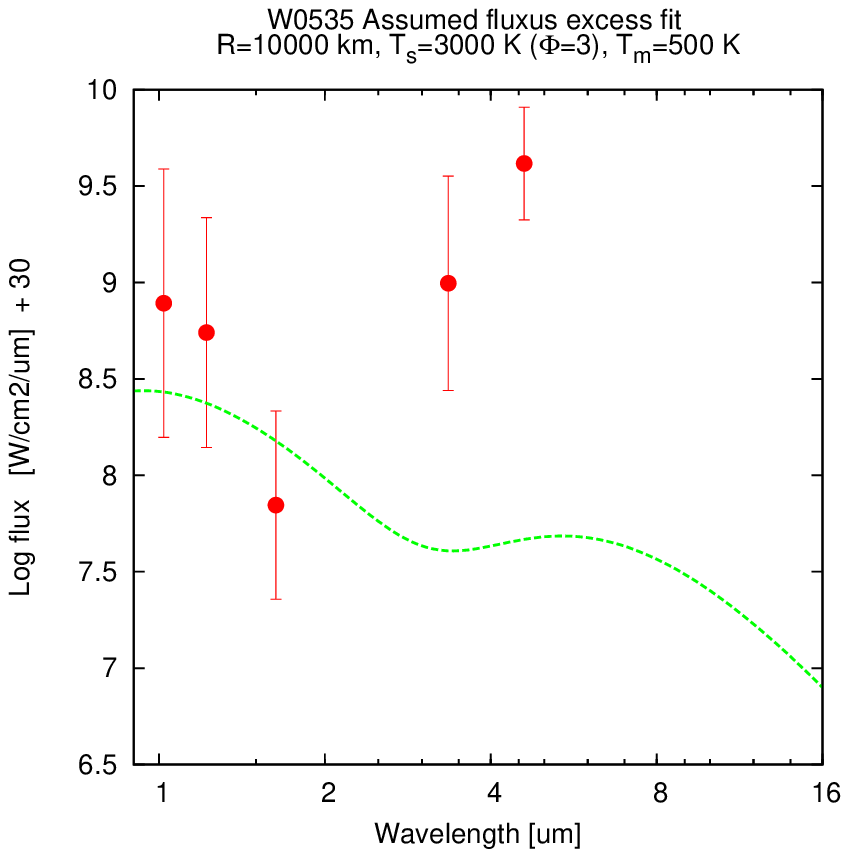,width=0.45\textwidth}
\psfig{figure=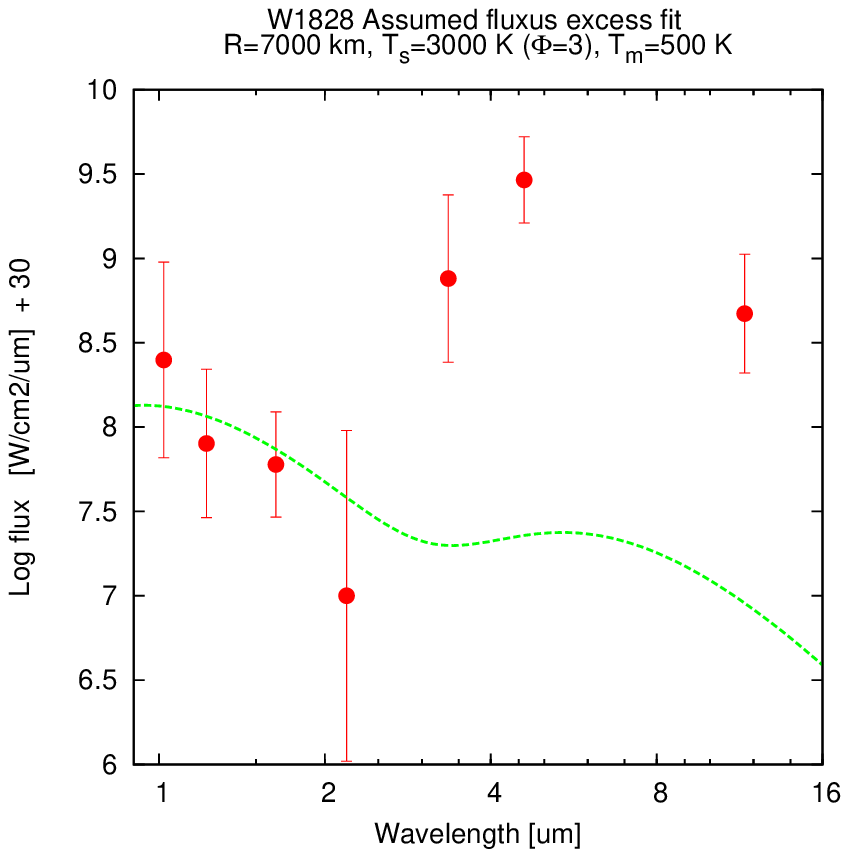,width=0.45\textwidth}
\caption{ Fitting the excess coming from the H-W2 (Fig. \ref{fits1}) and J-W2 color magnitude diagrams with tidal heated spotted moons.
}
\label{fits3}
\end{center}
\end{figure*}

However the assumed model (Figs. \ref{fits3})
cannot fit the W1--W2--W3 bands. 
This part of the spectrum requires a large and warm body, 
possibly a second BD companion. This conclusion is in accord with \citep{Leggett-2017-Bdwarf}, 
and show the limitations of our simplistic approach. 
The complete SED modeling of these enigmatic BD stars would require 
a detailed modeling of the non-thermal features -- which is the dominant in BD spectra -- and probably, 
new measurements with high S/N ratio. This is beyond the scope of this paper. 
Here we only attempted to give an idea for the origin of the hot part (Y--J--H bands) of the unexplained infrared excess, 
suggesting the presence of a tidally heated companion with a hot spot in these BD systems.



\section{\bf Conclusions and Discussion}

We created a non homogeneous model for the tidal heating of moons around exoplanets, where all the energy of the tidal heating
is radiated from a hot spot.
This can increase the bolometric flux of the moon with a factor of about a 2.5 times and up to 4 times, or 1-1.5 magnitudes.
If the spot is small, then the spot temperature becomes high,
and the energy emission is shifted to shorter wavelengths. 
Larger spots have smaller increase in the flux, but their spectrum will be better separated from that of the star, due to the lesser increasing effective temperature of the moon.


Super Earth companions around red dwarf stars have a high change to be detectable at a precision level of 100$+$ ppm. This precision can be realised in the future for many stars. Around 8 $\mu$m, Ariel has a S/N ratio better than 100 ppm in 1 hour for M0 stars within 30 pc, and for Solar-type stars until 50 pc following the same strategy (AIRS bands between 6--8 $\mu$m averaged). The accuracy in three elements covering the 7--8 $\mu$m range is better than 100 ppm for a K0 star not farther than 30 pc, and for an F0 star, the 100 ppm$\sqrt{h}$ precision can be reached in the single element at the 8 $\mu$m end. JWST between 5-30 $\mu$m has also a better S/N ratio than 100 ppm in 1 hour for M2 stars within 300 pc, and for G2 stars within 1000 pc in 1.5 hour.
\footnote{https://jwst.etc.stsci.edu/}

The spot temperature is around 500-3000 K (depending on the spot size) which we also consider to be realistic.
Smaller moons may also have a change for detection ($\sim$ 10 -- 30 ppm) if they are closer to the star.
Around sun like stars, even super earth sized, tidally heated moons have a very low contribution to be detected stellar flux ($\sim$ 10 ppm).


The detectability of the moon can also expressed in terms of the moon-to-planet flux ratio.
The planet flux itself can even be smaller then the flux of the moon if the planet and the moon are far from the star (>1 AU), and the eccentricity is large enough. The absolute brightness of these moons can reach 21 AB mag, 24 bol mag (super earths),
23 AB mag, 26 bol mag (earths), 25 AB mag, 28 bol mag (Ganymedes).
We considered that rouge planets (planets without stars) can be detectable due to their tidal heated moons.

This model is simple, and the realistic cases may include multiple different sized spots with different temperatures, resulting more complex spectrum.



We made simulations using Monte-Carlo to see that, within realistic bonds, what flux they give.
The primary parameter is the radius, bigger moons means bigger flux, 
however the density does not have significant effect, but bigger the better.
The second important parameter is the orbital period, smaller orbital period means bigger fluxes.
The eccentricity is not that important, while bigger eccentricity means bigger fluxes, it has a maximum around 0.21.
Bigger planet can host bigger moons, and planets with higher density are also better.\\

Super Earth moons around M2 stars can (Orbital period 1 day, eccentricity 0.1) can reach 100+ ppm compared to the star (if the spot radius between 20 and 120 degree).
Smaller moon still has 30 ppm if they closer to the star or they show activity.
The brightness of super Earths are 21 AB absolute magnitude (at 10 $\mu$m) and 24 bolometric absolute magnitude,
23 AB absolute magnitude (at 10 $\mu$m) and 26 bolometric absolute magnitude for Earths and 
25 AB absolute magnitude (at 10 $\mu$m) and 28 bolometric absolute magnitude for Ganymedes.

If the hot spot is not too small (and thus too hot), then super Earth moons (if they exists) around M2 star has a high change to be detected (as \citealt{Forgan-2017} suggested).
Moon far away from stars when they can be separated or rogue planets, 
has higher changes, where even Earths or Ganymede sized moons are detectable, because they are brighter then planet.\\

On Io, Pele is by far the most prominent hot spot, but there are at least several others \citep{McEwen-2001}.
So while multiple hot spots would be more realistic, a single hot spot case is a good first approximation.
However the multi hot spot scenario would be very complicated, because it involves calculations of internal heat transfer; while the parameter space increases by $\sim$4 parameters. These parameters include the new spot size, its location around the surface, which depends on the internal structure, and the distribution of the tidal heat between the spots. The physically plausible configuration of all of these parameters are an unexplored question.
The presence of the second spot smoothes the signal, because of two reasons. First, the heat is redistributed between 2 spots, so both are cooler than one single spot would be. And secondly, the two spots are farther or closer apart from each other, hence, the photometric effects occur at different times. This smearing reduces the full amplitude.\\

We also note that the assumption of synchronous rotation is used in our calculations. Synchronicity is, however, very plausible here, because the time scale of tidal locking of tidally heated moons is very short \citep{Dobos-survival}. These moons orbit close to the planet, the energy dissipation is very significant and the system evolves toward the most favourable state, which is the tidal locking. In the heating power of the tides, the most sensitive parameter is the eccentricity, which also suffers tidal decay (leading to circularisation of the orbit), but on the other hand, some eccentricity has to survive stable to support enough internal shear to heat up the body. Here a Kozai-like \citep{Kozai-1962} process is generally assumed (without giving details of how it acts on the actual system), and hence, a momentary non-zero value of the slowly oscillating eccentricity is plausible. This value is always introduced as a fixed parameter (or has been so in the previous papers on this topic at least), which does not require to dig into the dynamics in details.

In our paper, these are not serious problems, because we do not account for systems under a dynamical evolution at all. Our question is which is the part of the parameter space where there is enough amplification in respect to the already known tidal heating models to boost the signal from a inhomogeneous moon surface considerably. Discussing whether these parameters are likely or not in any stage of the dynamical evolution is out of the scope of this paper about this question.

\section{\bf Summary}

In this paper, we examined the observable effects of a hot spot on a tidally heated moon, and get the following main results.

\begin{enumerate}
    \item The amplification factor that describes how the thermal signal of the moon is increased by the spot is always $<4$, and is a monotonical decreasing function of the spot size in the bolomertic range. In a certain band, the greatest amplification occurs at a preferred non-zero optimal spot size which depends on the power of tidal heating, and through this, the geometrical configuration.
    
    \item{} The most dominant parameters that determine the detectability are: the size of the moon, the eccentricity of the moon and the orbital period. Unlike the size and the period, there is a non-monotonical dependence upon the eccentricity because of dynamical constraints: the moon has to avoid a impact or break-up at the Roche limit. 
    
    \item{} We described the tidally heated moons around the theorized rouge planets that do not orbit a host star. If these have a moon and considerably tidal heat, there is chance that the moon itself overshines the planet and the moon will be the major observable component of these systems, and their absolute magnitudes can reach up to 21 AB mags at MID and near infrared wavelengths. We also examined the known brown dwarfs with unusual spectra for the presence of a tidally heated moon, but did not get conclusive results from already available data.

    \item{} The effects of a tidally heated, spoted moon around a planet is in the order or less than 100 ppm at MID infrared wavelengths. Therefore, there is hope for the detection of these spots in the close future with instruments such as Ariel (up to 30--50 pc) or JWST (up to 100--500 pc).
    
    \item{} Small spot size (<10$^{\circ}$) leads to very low (<1) ppm values (Moon-to-star flux ratio) in MID infrared wavelengths even with super Earths moons, which makes them very hard to find.
    Because of this, small spot sized moons are more likely be found around rogue planets or brown dwarf, as probably they overshine their host.

\end{enumerate}

\section*{Acknowledgments}

This project enjoyed the generous support of the City of Szombathely under agreement No. S-11-10.

This work is part of the project
``Transient  Astrophysical  Objects``  GINOP  2.3.2-15-2016-00033 of the National Research, Development and
Innovation  Office  (NKFIH),  Hungary,  funded  by  the European  Union.

We thank Vera Dobos for the thorough reading of the manuscript and the detailed discussions about the
various tidal heating mechanisms

We also thank Bálint Süle for the detailed discussions about the Nusselt number and its implications.\\


\section*{Data availability}

The data underlying this article are available in
\url{https://github.com/Hydralisk24/Science}

\end{document}